\documentclass[aps,superscriptaddress,pra,twocolumn,showpacs,10pt]{revtex4-1}
\usepackage[english]{babel}
\usepackage{mathtools}
\usepackage{graphicx}
\usepackage{amsmath}
\usepackage{amsfonts}
\usepackage{amstext}
\usepackage{siunitx}
\usepackage{widetable}
\usepackage{xcolor}
\newcommand{\Tr}{\mathop{\mathrm{Tr}}\nolimits}
\newcommand{\ket}[1]{\left|#1\right\rangle}
\newcommand{\bra}[1]{\left\langle#1\right|}
\newcommand{\braket}[2]{\left\langle#1\middle|#2\right\rangle}
\newcommand{\epsmean}{\varepsilon_{\text{avg}}}
\newcommand{\rmd}{\mathrm{d}}
\newcommand{\rmi}{\mathrm{i}}
\newcommand{\rme}{\mathrm{e}}
\newcommand{\beall}[1]{\ket{\psi_{\rm in}^{(#1)}}}
\newcommand{\kiall}[1]{\ket{\psi_{\rm out}}^{(\rm #1)}}
\newcommand{\kozall}[1]{\ket{\psi_{\rm mid}^{(#1)}}}
\begin{document}
\title{Quantum state engineering via coherent-state superpositions in traveling optical fields}
\author{Emese Molnar}
\affiliation{Institute of Physics, University of P\'ecs, H-7624 P\'ecs, Ifj\'us\'ag \'utja 6, Hungary}
\author{Peter Adam}
\affiliation{Institute of Physics, University of P\'ecs, H-7624 P\'ecs, Ifj\'us\'ag \'utja 6, Hungary}
\affiliation{Institute for Solid State Physics and Optics, Wigner Research Centre for Physics, Hungarian Academy of Sciences, H-1121 Budapest, Konkoly-Thege Mikl\'os \'ut 29-33., Hungary}
\author{Gabor Mogyorosi}
\affiliation{Institute of Physics, University of P\'ecs, H-7624 P\'ecs, Ifj\'us\'ag \'utja 6, Hungary}
\author{Matyas Mechler}
\affiliation{MTA-PTE High-Field Terahertz Research Group, H-7624 P\'ecs, Ifj\'us\'ag \'utja 6, Hungary}
\date{\today}
\begin{abstract}
We propose two experimental schemes for producing coherent-state superpositions which approximate different nonclassical states conditionally in traveling optical fields. Although these setups are constructed of a small number of linear optical elements and homodyne measurements, they can be used to generate various photon number superpositions in which the number of constituent states can be higher than the number of measurements in the schemes. We determine numerically the parameters to achieve maximal fidelity of the preparation for a large variety of nonclassical states, such as amplitude squeezed states, squeezed number states, binomial states and various photon number superpositions. The proposed setups can generate these states with high fidelities and with success probabilities that can be promising for practical applications.
\end{abstract}
\maketitle
\section{Introduction}
Generation of various nonclassical states of light is still an important topic in quantum optics, owing to the numerous applications of such states in quantum information processing, quantum-enhanced metrology, and fundamental tests of quantum mechanics. The preparation of states in traveling optical modes is generally desired in many practical applications. Conditional preparation is a well-established technique for this task. This consists in measuring one of the modes of a bipartite correlated state which results in the projection of the other mode to the desired state for certain results of the measurement. Though this technique is probabilistic and generally approximative, it can provide quantum states with high enough fidelity for practical use.

Special attention has been devoted to the generation of coherent-state superpositions referred to as Schr\"odinger cat states in traveling modes \cite{jeong2004, lund2004, JeongPRA2006, LancePRA2006, NielsenPRA2007, glancy2008, MarekPRA2008, TakeokaPRA2008, he2009, LeePRA2012, LaghaoutPRA2013, wang2013} due to their important role as basis states in optical quantum information processing \cite{jeong2001, vanenk2001, jeong2003, ralph2003, AsbothEPJD2004, lund2008}. The components of these superpositions are two macroscopically distinguishable coherent states with opposite phases.
These states have already been prepared in several traveling wave experiments \cite{neegard2006, ourjoumtsev2007, TakahashiPRL2008, GerritsPRA2010, HuangPRL2015}, however, further efforts are needed for producing Schr\"odinger cats with larger amplitudes and higher fidelity to meet the criteria of the developed applications.
 
Quantum state engineering has also been extensively studied with the general aim of the preparation of a variety of different nonclassical states in the same single experimental scheme \cite{szabo1996, dakna1999, fiurasek2005, GerryPRA2006, BimbardNP2010, LeePRA2010, SperlingPRA2014, adam2015, HuangPRA2016}. It is a plausible approach to construct systematically the photon number expansion of the quantum states up to a given photon number. For realizing this task various methods have been developed, such as repeated photon additions \cite{dakna1999}, photon subtractions \cite{fiurasek2005} and the application of the superpositions of these processes \cite{LeePRA2010, SperlingPRA2014}. It is a characteristic property of such schemes that the number of the optical elements is generally proportional to the amount of number states involved in the photon number expansion of the target state. This implicates that an increase in the number of the constituent photon number states of the target state leads to a decrease in the success probability and even to that in the fidelity of the generation.

The possibility to overcome this issue is offered by the idea of quantum state engineering via discrete coherent-state superpositions. It has been shown that superpositions of even a small number of coherent states placed along a straight line, on a circle or on a lattice in phase space can approximate nonclassical field states with a high degree of accuracy \cite{janszky1993, janszky1995, szabo1996, adam2015}. For certain quantum states the number of the required coherent states for an approximation with a given accuracy can be less than that of the terms of the number-state expansion of the target state. Interestingly, different superpositions of various geometries can approximate the same nonclassical state \cite{adam2015}. 
This feature can be explained by the overcompleteness of the coherent states as a basis and even of discrete subsets of them in the Hilbert space of a harmonic oscillator \cite{stergioulas1999}.
It is still an interesting open question how to find the smallest number of coherent states whose superposition approximates the desired state with a given precision. Intuitively one can state that the best superpositions consist of coherent states whose position and geometry in phase space  ``fit well''  to the Wigner function of the desired state~\cite{szabo1996, adam2015}.
Several methods have been proposed for generating discrete coherent-state superpositions on a circle or along a line in phase space for electromagnetic fields in cavities \cite{szabo1996, DomokosPRA1994, LutterbachPRA2000, PathakPRA2005} and for the center of mass motion of a trapped ion \cite{MoyaCessaPRA1999}. An experimental scheme has also been developed for generating Fock states in a single-mode traveling-wave optical field based on coherent-state superposition on a circle \cite{GerryPRA2006}. Apart from this latter paper, quantum state engineering of traveling-wave optical fields based on coherent-state superpositions appears to be a largely unexplored area.

In this paper we propose two experimental schemes containing only a small number of linear optical elements and homodyne measurements that can be used for producing coherent-state superpositions along a line and on a lattice in phase space. These superpositions can approximate various nonclassical states in traveling optical fields. The input states of the schemes are superpositions of two coherent states with small phase separation and additional squeezed vacuum states in one of the schemes. These coherent-state superpositions can be generated by the scheme containing cross-Kerr nonlinearities as described in Ref.~\cite{gerry1999}. The analysis of the performance of that scheme under decoherence shows that these states are practically realizable ones \cite{jeong2005} and the necessary phase shift can be achieved by weak cross-Kerr nonlinearities realizable in present experiments~\cite{MatsudaNP2009, HePRA2011, HoiPRL2013, BartkowiakJPB2014}. In our proposed schemes the nonclassical states are prepared conditionally depending on the results of the homodyne measurements. Our description of the schemes leads to an optimization problem to determine the optimal parameters of the homodyne measurements and the input coherent-state superpositions to yield the desired states. We have found that this can be solved efficiently with genetic algorithms. We demonstrate through a broad variety of examples that amplitude squeezed states, squeezed number states, binomial states and various photon number superpositions can be generated in the proposed schemes with a high precision and with sufficient probabilities for practical applications. An additional benefit of our schemes is that even though the number of the required measurements is fixed and small (2 or 3), they are capable of efficiently generating certain states containing a large number of nonzero coefficients in their photon number expansion.

The paper is organized as follows.
In Sec.~\ref{sec:2} we describe the schemes we propose and discuss how they can be applied for generating coherent-state superpositions.
In Sec.~\ref{sec:3} the generation of various nonclassical states is analyzed in details and actual examples are presented.
Finally, in Sec.~\ref{sec:concl} the results are summarized and conclusions are drawn.

\section{Conditional generation of coherent-state superpositions\label{sec:2}}

In this Section we present two schemes for generating various superpositions of a finite number of coherent states around the origin of the phase space. Both schemes are built from standard optical elements such as beam splitters and homodyne detectors and use light in experimentally feasible quantum states as inputs. As there is a broad variety of nonclassical states which can be well approximated by such superpositions, these schemes can generate light in quantum states close to these nonclassical states.

\begin{figure}[tb]	\centerline{\includegraphics[width=0.9\columnwidth]{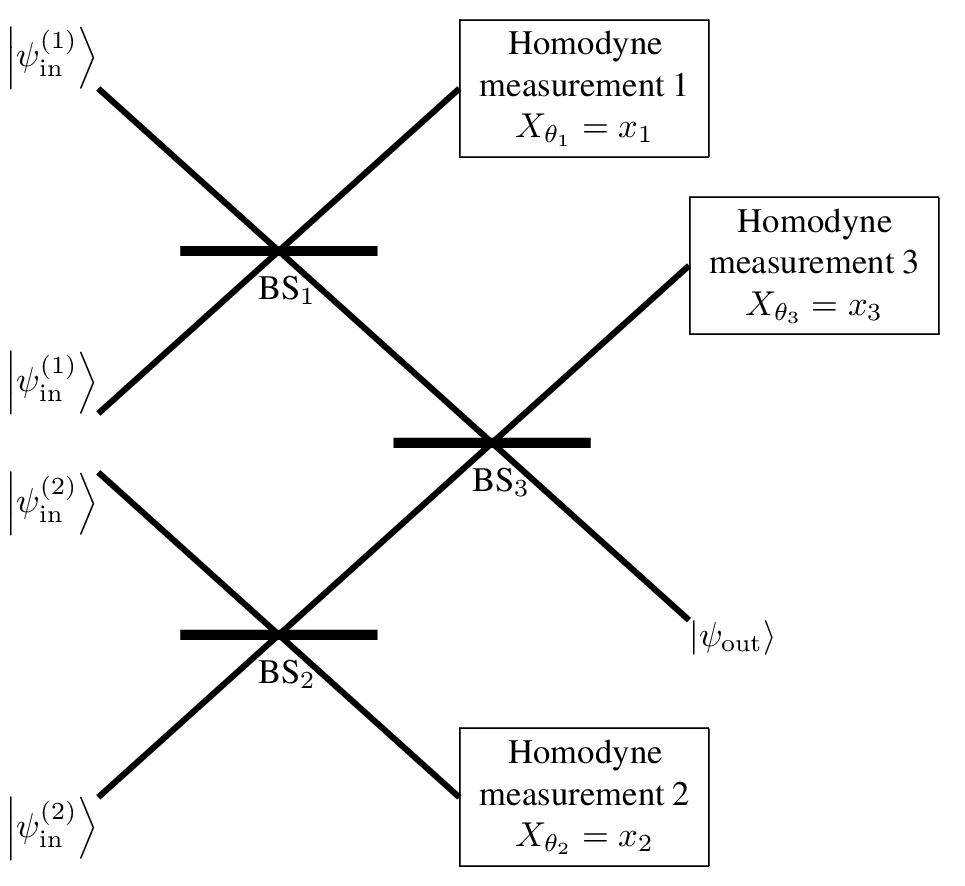}}
	\caption{Experimental scheme for generating superpositions of coherent states on a straight line or on a lattice. \label{fig:layout}}
\end{figure}

Let us first consider the scheme presented in Fig.~\ref{fig:layout}. The involved beam splitters are standard 50:50 ones.
We show that this setup is capable of producing coherent-state superpositions along a straight line or on a lattice in the phase space. The superpositions in argument are of the form
\begin{align}
	&\kiall{line} = \mathcal{N}_{\mathrm{line}} \cdot \sum_{n=-2}^2 c_n \ket{n \beta} , \label{eq:dcss_line}\\
	&\kiall{lattice} = \mathcal{N}_{\mathrm{lattice}} \cdot \sum_{k,l=-1}^1 c_{kl} \ket{k \beta+\rmi l\beta }. \label{eq:dcss}
\end{align}
The input states $\beall{i}$ of the system are assumed to be the following special coherent-state superpositions:
\begin{equation}
\beall{i} = \mathcal{N}_{\psi_{\mathrm{in}}} \cdot \left(\ket{\alpha_i} + \ket{\alpha_i\exp(-{\rmi}\varphi)} \right), \quad i=1,2, \label{eq:source}
\end{equation}
where $\mathcal{N}_{\psi_{\mathrm{in}}}$ is a normalization factor.
Such input states can be generated by experimental setups containing cross-Kerr nonlinearities \cite{gerry1999}.

The magnitudes of the coherent amplitudes $|\alpha_1|$ and $|\alpha_2|$ are the same in both of the states $\beall{1}$ and $\beall{2}$. The phase of the amplitudes, however,  can be different and depend on the desired output state.
When the superpositions of states along a straight line are to be generated, the input states have to be the same: $\beall{1}=\beall{2}$. In particular, if the superposition is required to be on the real axis of the phase space then the phase of the coherent amplitude in the initial superposition must be chosen as $\arg \alpha =\pi/2+\varphi/2$.
On the other hand, when the target superposition is taken on an orthogonal lattice oriented parallel to the phase space axes, the phase of the coherent amplitude in $\beall{1}$ is the same as in the previous case while the phase of the coherent amplitude in $\beall{2}$ must be chosen so that $\arg \alpha=\varphi/2$. With these choices for $\arg \alpha$ the parameter $\beta$ in \eqref{eq:dcss_line} and \eqref{eq:dcss} is real.

Homodyne measurements in the setup can measure the rotated quadrature operator $X_\theta$. The overlap between the eigenstate $\ket{x_\theta}$ of this operator and a general coherent state $\ket{\alpha}$ can be described by the inner product
\begin{eqnarray}
\braket{x_{\theta}}{\alpha} &=& \pi^{-\frac{1}{4}}\exp\left[-\frac12|\alpha|^2\right]\nonumber\\
&\times& \exp\left[-\frac{1}{2}x_\theta^2+\sqrt{2}\rme^{-\rmi \theta}x_\theta\alpha-\frac{1}{2}\alpha^2\rme^{-2\rmi\theta}\right].
\end{eqnarray}
Since we have already fixed the phase of the initial states the phase of the measured quadratures can be fixed as well.
Therefore we choose the phases of the homodyne measurements in Fig.~\ref{fig:layout} to be $\theta_1=\theta_2=\theta_3=0$ in the case of a superposition on a straight line, while these phases are chosen to be  $\theta_1=\theta_3=0$ and $\theta_2=\pi/2$ for the superposition on a lattice. In the latter case the second homodyne measurement measures the quadrature $Y$.

Using the well-known beam splitter transformations acting on coherent states as
\begin{equation}
	\ket{\alpha_1}_1 \otimes \ket{\alpha_2}_2 \rightarrow \ket{\frac{\alpha_1 + \alpha_2}{\sqrt{2}}}_3 \otimes \ket{\frac{\alpha_1 - \alpha_2}{\sqrt{2}}}_4 \label{eq:bs}
\end{equation}
and applying the projection $\ket{X_{\theta_i}=x_i}\bra{X_{\theta_i}=x_{i}}$ corresponding to the $i$th homodyne measurements ($i=1,2$) on one of the output modes of the beam splitters BS$_1$ and BS$_2$, we get the intermediate states $\kozall{1}$ and $\kozall{2}$ that serve as the input of the third beam splitter BS$_3$ in the form
\begin{eqnarray}
	\begin{array}{l}
		\kozall{1} = \mathcal{N}\left(a_0\ket{0} + a_1\ket{\text{cat}}\right), \\
		\kozall{2}^{(\rm line)} = \mathcal{N}\left(b_0\ket{0} + b_1\ket{\text{cat}}\right), \\
		\kozall{2}^{(\rm lattice)} = \mathcal{N}\left(b_0'\ket{0} + b_1'\ket{\text{cat}'}\right),
	\end{array}\label{eq:intermediate}
\end{eqnarray}
where the states $\ket{\text{cat}}$ and $\ket{\text{cat}'}$ are Schr\"odinger-cat states in the real and imaginary axis of phase space, respectively:
\begin{eqnarray}
	\begin{array}{l}
		\displaystyle 
		\ket{\text{cat}}=\ket{\sqrt{2}\alpha\sin\frac{\varphi}{2}}+\ket{-\sqrt{2}\alpha\sin\frac{\varphi}{2}},\\
		\displaystyle
		\ket{\text{cat}'}=\ket{\sqrt{2}\alpha\rmi\sin\frac{\varphi}{2}}+\ket{-\sqrt{2}\alpha\rmi\sin\frac{\varphi}{2}},
	\end{array}
\end{eqnarray}
and the coefficients $a_i$, $b_i$, and $b_i'$ take the form
\begin{eqnarray}
\begin{array}{rcl}
a_0 &=& \braket{x_1}{\sqrt{2}\alpha\rmi\rme^{\rmi\frac{\varphi}{2}}} + \braket{x_1}{\sqrt{2}\alpha\rmi\rme^{-\rmi\frac{\varphi}{2}}}, \\
a_1 &=& \braket{x_1}{\sqrt{2}\alpha\rmi\cos\frac{\varphi}{2}}, \\
b_0 &=& \braket{x_2}{\sqrt{2}\alpha\rmi\rme^{\rmi\frac{\varphi}{2}}} + \braket{x_2}{\sqrt{2}\alpha\rmi\rme^{-\rmi\frac{\varphi}{2}}}, \\
b_1 &=& \braket{x_2}{\sqrt{2}\alpha\rmi\cos\frac{\varphi}{2}}, \\
b'_0 &=& \braket{y_2}{\sqrt{2}\alpha\rme^{\rmi\frac{\varphi}{2}}} + \braket{y_2}{\sqrt{2}\alpha\rme^{-\rmi\frac{\varphi}{2}}}, \\
b'_1 &=& \braket{y_2}{\sqrt{2}\alpha\cos\frac{\varphi}{2}}.
\end{array}
\label{eq:abcoeffs}
\end{eqnarray}
From the latter two equations on, in the rest of this paper $\alpha$ denotes the magnitude (i.e. a real number, not including the phase) of the coherent state appearing originally in the input state defined in Eq.~\eqref{eq:source}. We note that in Eq.~\eqref{eq:intermediate} the coefficients of the vacuum state can approach zero for certain parameters of the input states and the results of the homodyne measurement, so this part of the scheme is suitable for preparing Schr\"odinger cat states with large amplitude \cite{adam2010}.

Considering the intermediate states described in Eq.~\eqref{eq:intermediate} it is easy to see that after the third homodyne measurement the output states are the ones given in Eqs.~\eqref{eq:dcss_line} and \eqref{eq:dcss} with the coefficients 
\begin{eqnarray}
c_{-2} &=& a_1b_1\braket{x_3}{0},\nonumber\\
c_{-1} &=& a_0b_1\braket{x_3}{\beta} + a_1b_0\braket{x_3}{-\beta}, \nonumber\\
c_0 &=& a_0b_0\braket{x_3}{0}+a_1b_1\braket{x_3}{2\beta}+ a_1b_1\braket{x_3}{-2\beta},\nonumber\\
c_1 &=& a_0b_1\braket{x_3}{-\beta} + a_1b_0\braket{x_3}{\beta}, \\
c_2 &=& a_1b_1\braket{x_3}{0}\nonumber
\label{eq:line_coeffs}
\end{eqnarray}
for the superposition along a line, and
\begin{eqnarray}
\begin{array}{rcl}
c_{-1,1} &=& a_1b_1'\braket{x_3}{-\beta-\rmi \beta} \\
c_{0,1} &=& a_0b_1'\braket{x_3}{-\rmi \beta}, \\
c_{1,1} &=& a_1b_1'\braket{x_3}{\beta-\rmi \beta}, \\
c_{-1,0} &=& a_1b_0'\braket{x_3}{-\beta}, \\
c_{0,0} &=& a_0b_0'\braket{x_3}{\beta}, \\
c_{1,0} &=& a_1b_0'\braket{x_3}{\beta}, \\
c_{-1,-1} &=& a_1b_1'\braket{x_3}{-\beta+\rmi \beta}, \\
c_{0,-1} &=& a_0b_1'\braket{x_3}{\rmi \beta}, \\
c_{1,-1} &=& a_1b_1'\braket{x_3}{\beta+\rmi \beta}
	\end{array}
	\label{eq:racs_coeffs}
\end{eqnarray}
for the superposition on a lattice. In these coefficients the coherent
amplitude reads
\begin{equation}
\beta = \alpha\sin(\varphi/2).\label{eq:beta}
\end{equation}
Note that this amplitude is identical to the one appearing in the superpositions of Eqs.~\eqref{eq:dcss_line} and \eqref{eq:dcss}.

\begin{figure}[tb]
\centerline{\includegraphics[width=0.9\columnwidth]{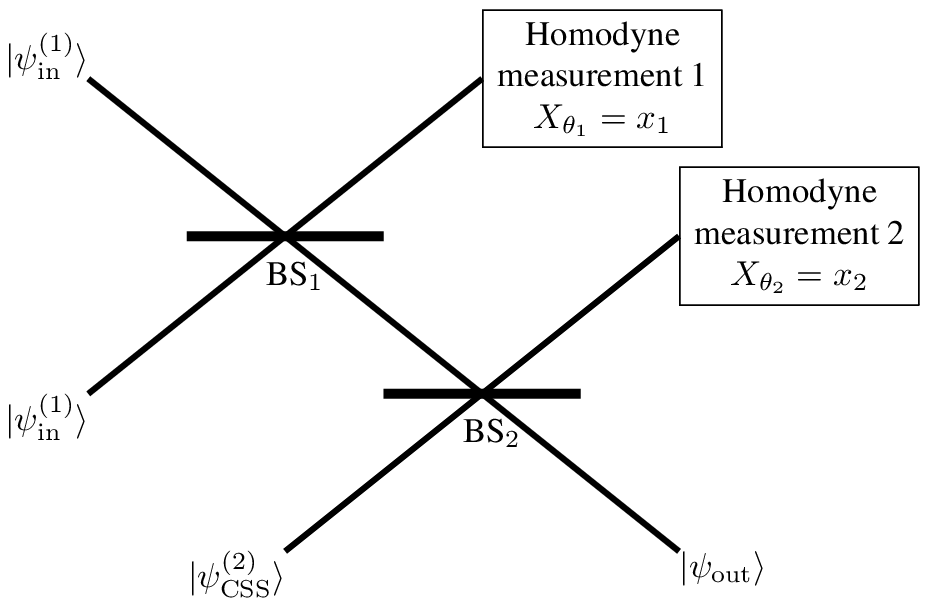}}
\caption{Experimental scheme with two homodyne measurements for generating superpositions of coherent states on a straight line or on a lattice.\label{fig:layout_5}}
\end{figure}

In Fig.~\ref{fig:layout_5} we propose a different scheme in which we replace one of the units producing the intermediate states of Eq.~\eqref{eq:intermediate} with an input state $\ket{\psi^{(2)}_\textrm{CSS}}$ which is the equidistant superposition of several coherent states along a line in phase space.  From a practical point of view this state can be a squeezed vacuum state $\ket{\zeta,0}$ with complex squeezing parameter $\zeta=r\exp(\rmi\theta)$. Such states can be approximated by coherent superpositions of a few coherent states of the form
\begin{equation}
\ket{\psi^{(2)}_\textrm{CSS}}=\sum_{l=-n}^{n} c'_{l}\ket{l\gamma\rme^{\rmi\frac{\theta}{2}}},\quad n=\frac{N-1}{2}\label{eq:psicss}
\end{equation}
with a high precision \cite{szabo1996}. In this equation the coherent amplitude $\gamma$ is real. For a real squeezing parameter $\zeta=r$ corresponding to squeezing in the variance of the $Y$ quadrature the coherent states in Eq.~\eqref{eq:psicss} are located along the real axis $x$. For complex squeezing parameter with $\theta=\pi$ these states are located along the imaginary axis $y$.

The coefficients $c'_l$ and the coherent amplitude $\gamma$ can be determined by using the method of Ref.~\cite{szabo1996}. The coefficients $c'_l$ can be derived from the one-dimensional coherent-state representation of the squeezed vacuum states and they read
\begin{eqnarray}
	c'_l = \mathcal{N} \cdot \exp\left(-\frac{1}{\rme^{2r}-1}|l\gamma|^2\right).
\end{eqnarray}
The value of the coherent amplitude $\gamma$ can be derived from the condition that the misfit between the approximating coherent-state superposition $\ket{\psi^{(2)}_\text{CSS}}$ and the original squeezed vacuum state $\ket{\zeta,0}$ should be minimal. The misfit of a target state and an approximate state is quantified in general by the parameter
\begin{equation}
\varepsilon=1-\left|\braket{\psi_\text{appr.}}{\psi_\text{target}}\right|^2,\label{eq:misfit}
\end{equation}
based on the fidelity $\left|\braket{\psi_{\text{appr.}}}{\psi_{\text{target}}}\right|^2$ of the states. In the actual setting $\ket{\psi_\text{target}}=\ket{\zeta,0}$ and  $\ket{\psi_\text{appr.}} =  \ket{\psi^{(2)}_\textrm{CSS}}$, but we shall use this quantification in all the other cases studied in this paper.

\begin{figure}
\centerline{\includegraphics[width=0.95\columnwidth]{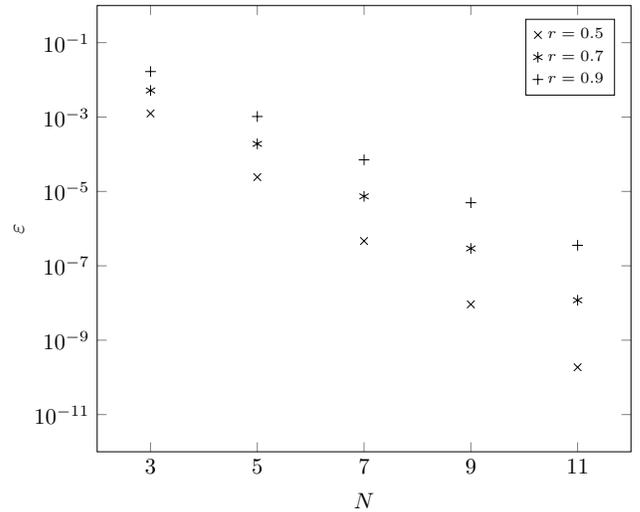}}
\caption{\label{MG_Fig3}The minimized misfit between the approximating coherent-state superposition and the original squeezed vacuum state as a function of the number of constituent states $N$ in the superposition for different values of the real squeezing parameter $r$.}
\end{figure}
In Fig.~\ref{MG_Fig3} we present the minimized misfit as a function of the number of constituent states $N$ in the superposition for different values of the squeezing parameter $\zeta$. This figure clearly shows that by increasing the number of constituent states the accuracy of the approximation also increases. The number of the required constituent states increases at a given accuracy by increasing the real squeezing parameters.

By using the superposition described in \eqref{eq:psicss} and the input state $\beall{1}$ given by \eqref{eq:source} with the same phase parameter used in the first setup, it is easy to see that the output of this second setup is the following general coherent-state superposition:
\begin{eqnarray}
	\ket{\psi_\text{out}} = \mathcal{N}_{\text{out}} \cdot \sum_{k=-1}^1\sum_{l=-n}^nc_{kl}\ket{k\beta - \frac{1}{\sqrt{2}}l\gamma \rme^{\frac{\rmi\theta}{2}}},\label{eq:scheme2out}
\end{eqnarray}
where
\begin{eqnarray}
	c_{kl} = a_{|k|} \cdot c'_l \cdot \braket{x_2}{k\beta + \frac{1}{\sqrt{2}}l\gamma\rme^{\frac{\rmi\theta}{2}}}.\label{eq:Nracscoeffs}
\end{eqnarray}
The coefficients $a_{|k|}$ in this expression are defined in Eq.~\eqref{eq:abcoeffs}.
These formulae describe various superpositions of $N\times 3$ coherent states depending on the value of the phase parameter $\theta$. For $\theta = 0$ one obtains superpositions along the real axis while for $\theta=\pi$ on an orthogonal lattice in the phase space.
In the latter case the lattice has $N$ elements in the direction of the imaginary axis and three elements in the other direction of the space.

We note here that in the examples presented in the next section, instead of an exact squeezed vacuum state as input for the scheme of Fig.~\ref{fig:layout_5}, we have calculated with approximating coherent-state superpositions containing $N=7$ coherent states. This choice results in misfits of $\varepsilon<10^{-4}$ for the required squeezed vacuum states occurring in the examples.

In order to use the schemes presented in Figs.~\ref{fig:layout} and \ref{fig:layout_5} for preparing a nonclassical state, one can apply the following strategy. First, a coherent-state superposition from Eqs.~\eqref{eq:dcss_line}, \eqref{eq:dcss}, and \eqref{eq:scheme2out} must be chosen for approximating a given target state. Next, all the parameters appearing in the chosen superposition must be determined in a way that the misfit $\varepsilon$ defined in Eq.~\eqref{eq:misfit} should be minimal between the target state and the approximating coherent-state superposition.

The variable parameters of the optimization include the measurement results $x_1$, $x_2$ or $y_2$, and $x_3$ of the homodyne measurements, the coherent amplitude $\alpha$ and the phase shift $\varphi$ of the input states $\beall{i}$ for the first setup.
For the second setup the corresponding parameters are the measurement results $x_1$ and $x_2$, the coherent amplitude $\alpha$ and the phase shift $\varphi$ of the input states $\beall{1}$, the squeezing parameter $r$ and the coherent amplitude $\gamma$ characterizing the other input $\ket{\psi_{\text{CSS}}^{(2)}}$. The optimization problem is neither linear nor convex, moreover, the relevant range of the parameters depend also on the physical circumstances. In spite of these difficulties we have found that it can be efficiently solved e.g.\ using genetic algorithms.

Finally, let us introduce the other figures of merit, in addition to the misfit, which are commonly used to characterize the performance of conditional quantum state generating schemes. In the case of conditional schemes for generating field states of a traveling mode the probability of success is also important. For homodyne measurements the probability of obtaining precisely an eigenvalue of the given quadrature operator is obviously zero as the quadratures are continuous variables. Hence, probability of success corresponding to a single measurement event resulting $x_i^{\rm opt}$ by the $i$th homodyne measurement after the  beam splitter BS$_i$ in the schemes of Figs.~\ref{fig:layout} and \ref{fig:layout_5} are to be defined as \cite{jeong2004}
\begin{eqnarray}
P^{(i)}\left(x_i^{\rm opt},\delta_i\right)=\intop_{x_i^{\rm opt}-\delta_i}^{x_i^{\rm opt}+\delta_i}\rmd x_i\Tr\left[\rho^{(i)}_3\ket{x_i}\bra{x_i}\right]
	\label{eq:prob1number}
\end{eqnarray}
where 
\begin{equation}
\rho^{(i)}_3=\Tr_4\left[\ket{\psi^{(i)}}_{34} \prescript{}{34}{\bra{\psi^{(i)}}}\right]
\end{equation}
is the  density operator of the mode on which the $i$th homodyne measurement is performed. The state $\ket{\psi^{(i)}}_{34}$ is the two-mode state after the $i$th beam splitter. The explicit form of these states can be obtained via a straightforward calculation from Eqs.~\eqref{eq:source} and \eqref{eq:bs}. The quantity $\delta_i$ defines the range in which the misfit parameter $\varepsilon$ in Eq.~\eqref{eq:misfit} is smaller than a prescribed value. We define the overall probability of success $P$ as the product of the success probabilities $P^{(i)}$:
\begin{equation}
P=\prod_i P^{(i)}\left(x_i^{\rm opt},\delta_i\right).\label{eq:overallP}
\end{equation}
We note that in the first scheme the state  $\ket{\psi^{(3)}}_{34}$ depends on the results of the previous measurements, while in the second scheme the state $\ket{\psi^{(2)}}_{34}$ depends on the result of the first homodyne measurement.
If the measurement ranges are small enough the success probability of the final measurement can be calculated at the optimal parameter values of the previous measurements and the independence of the individual measurements assumed in Eq.~\eqref{eq:overallP} can be considered to be valid.

Note that the misfit parameter changes with the measurement results within the measurement ranges. Therefore it is interesting how one can characterize the accuracy of the preparation of the given state after choosing certain measurement ranges for the particular homodyne measurements. Dividing all ranges to sufficiently small subranges we can assign a certain measurement results for each of the subranges. Next, we can calculate the misfit parameter and the overall probability of success by the application of the formulas introduced earlier for any of the possible measurement outcomes, that is, for any particular combination of the subranges. Finally, we can introduce the average misfit of the preparation as
\begin{equation}
	\epsmean = \frac{\sum_j P_j\cdot\varepsilon_j}{\sum_j P_j},
\end{equation}
where $P_j$ and $\varepsilon_j$ are the overall probability of success and the misfit of the $j$th outcome.

\section{Examples of generating nonclassical states \label{sec:3}}
Thus far we have presented schemes which are capable of generating superpositions of coherent states. In this Section, we demonstrate through examples how they can be applied to generate a large variety of nonclassical states. Our examples include  amplitude squeezed states, binomial states, squeezed number states, and special photon number state superpositions that can be important for protocols used in quantum optics and quantum information science \cite{MarekPRA2009}. Let us first recapitulate the definitions and some properties of the nonclassical states whose generation we address.

Amplitude squeezed states are defined by Gaussian continuous coherent-state superpositions on circles in phase space:
\begin{equation}
	\ket{\alpha_0,u,\delta}_{\text{AS}} = c\int\exp\left(-\frac{1}{2}u^2\phi^2-\rmi\delta\phi\right)\ket{\alpha_0\rme^{\rmi\phi}}{\rm d}\phi,
	\label{eq:banan}
\end{equation}
where $c$ is a normalization constant, $u$ determines the width of the distribution, $\alpha_0$ is the magnitude of the amplitudes of the superposed coherent states and $\delta$ is a free modulation constant. For $u\rightarrow \infty$ the distribution contracts into the coherent state $\ket{\alpha_0}$. In the opposite limit $u \ll 1$ one has the $n$-photon number state with $n=\delta$. The state $\ket{\alpha_0,u,\delta}_{\text{AS}}$ can be expanded in photon number basis as
\begin{eqnarray}
\ket{\alpha_0,u,\delta}_\text{AS} = c \sum_{n=0}^{\infty}\frac{\sqrt{2\pi} \alpha_0^n}{u\sqrt{n!}}\exp\left[-\frac{(\delta-n)^2}{2u^2}\right]\ket{n}.
	\label{eq:asstate_photon}
\end{eqnarray}
These states are intelligent states of the Pegg-Barnett number-phase uncertainty relation \cite{AdamPLA1991, AdamSzalayPRA2014}.

A single-mode binomial state can be defined as the following number-state expansion \cite{Stoler1985}:
\begin{equation}
\ket{p,M}_\text{B}=\sum_{n=0}^MB_n^M\ket{n},\label{eq:binom1}
\end{equation}
where the $B_n^M$ coefficients are
\begin{equation}
B_n^M=\left[\frac{M!}{n!(M-n)!}p^n(1-p)^{M-n}\right]^{1/2}.\label{eq:binom2}
\end{equation}

From Eqs.~\eqref{eq:binom1} and \eqref{eq:binom2} it can be seen that given any finite $M$, if $p=0$, $\ket{p,M}$ is reduced to the vacuum state $\ket{0}$. On the other hand, if $p=1$, we obtain the number state $\ket{n=M}$. In the limit $p\to0$ and $M\to\infty$, but with $pM=\alpha^2$ constant, $\ket{p,M}$ becomes a coherent state $\ket{\alpha}$.

The squeezed number states  \cite{kim1989,liu1993,kral1990} can be described by
\begin{eqnarray}
\ket{n,\zeta}_{\text{NS}}&=&\hat{S}(\zeta)\ket{n}\nonumber\\
&=&\sum_m\ket{m}\bra{m}S(\zeta)\ket{n}\nonumber\\
&=&\sum_m\ket{m}G_{mn}(\zeta),
\end{eqnarray}
where $\hat{S}(\zeta)$ is the squeezing operator. We do not recapitulate the explicit formula for the coefficients $G_{mn}$ here, for its length. It can be found e.g.\ in Eq.~(34) of Ref.~\cite{PRD1985}. It is clear that all the introduced states can contain several photon number states in their number-state expansions with non-negligible coefficients for certain parameters. So they can be used to demonstrate that the proposed schemes can generate photon number superpositions containing far more constituent states than the number of measurements in the proposed schemes.

Finally, we will consider the generation of the following particular photon number superpositions containing photon number states of a few photon numbers:
\begin{eqnarray}
\begin{array}{rc>{\displaystyle}l}
 \ket{\psi_{02}} &=&  \frac{1}{\sqrt{10}}\left(3\ket{0}+\ket{2} \right),\\
\ket{\psi_{02}}' &=& \frac{1}{\sqrt{2}}(\ket{0}+\ket{2}),\\
 \ket{\psi_{012}} &=& \frac{1}{\sqrt{18}}\left( 4\ket{0}+\ket{1}+\ket{2}\right),\\
\ket{\psi_{012}}'&=& \frac{1}{\sqrt{6}}(2\ket{0}+\ket{1}+\ket{2}),\\
\ket{\psi_{012}}'' &=& \frac{1}{3}(2\ket{0}+2\ket{1}+\ket{2}),\\
\ket{\psi_{0123}} &=& \frac{1}{\sqrt{99}}\left( 8\ket{0}+5\ket{1}+3\ket{2}+\ket{3}\right).
\end{array}
\end{eqnarray}
In these superpositions the coefficients are chosen in an ad hoc manner.

In our calculations we have chosen to use a genetic algorithm \cite{goldberg} for solving the optimization problem. In order to minimize the misfit of the approximating state we have to define the ranges of the optimization parameters so that the bounds are physically reasonable and the optimization problem is numerically stable and feasible.
First the ranges of the measurement parameters $x_i$ have to be chosen so that the coefficients in the superpositions \eqref{eq:dcss_line}, \eqref{eq:dcss}, and \eqref{eq:scheme2out} depending on these parameters can take all their possible values. We found that in all the considered examples the values of the measurement parameters took their values within the range $-10$ to $10$, therefore we have set the range of the optimization for these parameters accordingly. Similarly, in the scheme of Fig.~\ref{fig:layout_5} we chose the range of the squeezing parameter by broadening it empirically to make all the desired approximate states feasible.

\begin{table}[tb]
\caption{The minimal misfit $\varepsilon$, the corresponding input-state parameters $\alpha$, $\varphi$, and the measurement results $x_1$, $x_2$, and $x_3$ of the homodyne measurements for the amplitude squeezed state $\ket{1,2,1}_{\text{AS}}$ approximated on a lattice using the scheme of Fig.~\ref{fig:layout_5}. The optimization was performed for different ranges of the coherent amplitude $\alpha$ and the phase difference $\varphi$.\label{tab:racs2_1}}
\begin{tabular*}{\columnwidth}{
S[table-format=1.3e1,table-sign-exponent]@{\extracolsep{\fill}}
S[table-format=5.0]
S[table-format=1.1e1,table-sign-exponent]
S[table-format=1.2]
S[table-format=1.3, table-sign-mantissa]
S[table-format=1.2,table-sign-mantissa]
S[table-format=1.3]}
\toprule
{$\varepsilon$}&{$\alpha$}&{$\varphi$}&{$\beta$} & {$x_1$} &{$x_2$} &{$x_3$}\\ \hline\\[-8pt]
1.421e-4 & 22981 & 7.1e-5 & 0.82 & 2.26 & -2.18 & 4.13\\
1.031e-4 & 4363 & 3.9e-4 & 0.85 & 2.039 & -2.2 & 3.81\\
1.027e-4 & 349 & 4.9e-3 & 0.86 & 1.97 & -2.18 & 3.75\\
1.137e-4 & 886 & 2e-3 & 0.89 & 1.83 & 2.15 & 3.54\\ 
1.046e-4 & 347 & 5.1e-3 & 0.87 & -1.91 & 2.17 & 3.65\\
1.055e-4 & 238 & 7.3e-3 & 0.88 & 1.88 & -2.16 & 3.62\\
1.586e-4 & 698 & 2.3e-3 & 0.8 & 2.35 & -2.19 & 4.241\\
1.046e-4 & 417 & 4.2e-3 & 0.88 & -1.89 & 2.17 & 3.62\\
1.106e-4 & 307 & 5.5e-3 & 0.85 & 2.08 & 2.2 & 3.87\\
\botrule
\end{tabular*}
\end{table}

\begin{figure}[tb]
\includegraphics[width=\columnwidth]{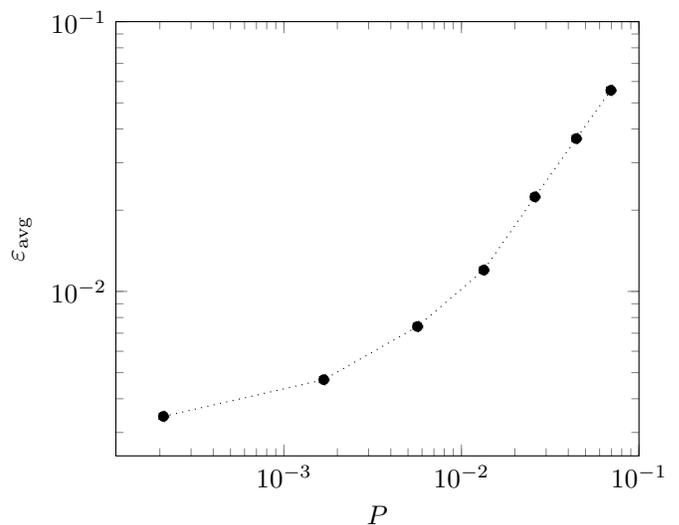}
\caption{\label{fig:eavgP}The average misfit $\epsmean$ as a function of the overall probability $P$ for the state $\ket{0.2,10}_{\rm B}$ approximated on a lattice in the scheme presented in Fig.~\ref{fig:layout}.}
\end{figure}

\begin{table*}[tb]
\caption{Results of the optimization for different nonclassical states for the scheme of Fig.~\ref{fig:layout}. The states are approximated by coherent-state superpositions along a line described by Eq.~\eqref{eq:dcss_line}. The table presents for each state the misfit $\varepsilon$ of the approximation and the optimized parameters leading to this misfit including the coherent amplitude $\alpha$, the phase distance $\varphi$, and the resulting coherent amplitude $\beta$ of the input states, the measurement results $x_i$ of the homodyne measurements, and the overall probability $P$, the corresponding ranges $\delta_i$ of the measurements, and the average misfit $\epsmean$.\label{tab:emese2}}
\begin{tabular*}{\textwidth}{
l@{\extracolsep{\fill}}
S[table-format=1.1e1,table-sign-exponent]
S[table-format=3.0]
S[table-format=1.1e1,table-sign-exponent]
S[table-format=1.2]
S[table-format=1.2,table-sign-mantissa]
S[table-format=1.2,table-sign-mantissa]
S[table-format=1.2,table-sign-mantissa]
S[table-format=1.2]
S[table-format=1.3]
S[table-format=1.3]
}
\toprule
{state}&{$\varepsilon$}&{$\alpha$}&{$\varphi$}&{$\beta$} & {$x_1$} & {$x_2$} & {$x_3$} & {$\delta$} & {$P$} & {$\epsmean$} \\\hline\\[-8pt]
$\ket{1,1.5,1}_{\text{AS}}$ & 8.8e-5 & 616&1.7e-3 & 0.55 & 1.14 & 0.82 & 2.63& 0.75 & 0.004 & 0.073 \\
$\ket{1,2,1}_{\text{AS}}$ & 2.6e-4 & 245&6.3e-3 & 0.77 & -1.97 & -0.25 & -1.94& 0.35 & 0.025 & 0.011 \\
$\ket{\sqrt{2},2.5,2}_{\text{AS}}$ & 7.5e-3 & 698 & 3.7e-3 & 1.29 & 2.13 & -1.03 & -1.52 & 0.3& 0.005 & 0.043 \\
$\ket{\sqrt{2},3,2}_{\text{AS}}$ & 4.4e-3 & 691 & 3.1e-3 & 1.31 & -2.17 & -1.0 & -1.61 & 0.4 & 0.012 & 0.068 \\
$\ket{0.1,5}_{\text{B}}$ & 2.5e-4 & 780 & 6.4e-3 & 0.6 & 0.29 & 3.72 & 2.62 & 1.0 & 0.004 & 0.041 \\
$\ket{0.3,6}_{\text{B}}$ & 9.6e-3 &492&5.2e-3 & 1.29 & -0.38 &2.52 &0.8& 0.4 & 0.014 & 0.024\\
$\ket{0.2,8}_{\text{B}}$ & 2.8e-3 &939&2.9e-3 & 1.39 & 0.54 &-1.86 &2.17& 0.5 & 0.017 & 0.024\\
$\ket{\psi_{012}}$ & 3.4e-4 & 269 & 1.9e-3 & 0.26 & -3.99 & 0.1 & -0.45 & 0.5 & 0.085 & 0.019 \\
$\ket{\psi_{0123}}$ & 2.0e-3 & 540 & 2.6e-3 & 0.71 & 3.44 & 1.63 & -3.59 & 1.5 & 0.005 & 0.077\\
\botrule
\end{tabular*}
\end{table*}
\begin{table*}[tb]
\caption{Results of the optimization for different nonclassical states for the scheme of Fig.~\ref{fig:layout}. The states are approximated by coherent-state superpositions on a lattice described by Eq.~\eqref{eq:dcss}. The table presents for each state the misfit $\varepsilon$ of the approximation and the optimized parameters leading to this misfit including the coherent amplitude $\alpha$, the phase distance $\varphi$, and the resulting coherent amplitude $\beta$ of the input states, the measurement results $x_i$ of the homodyne measurements, and the overall probability $P$, the corresponding ranges $\delta_i$ of the measurements, and the average misfit $\epsmean$.\label{tab:emese3}}
\begin{tabular*}{\textwidth}{
l@{\extracolsep{\fill}}
S[table-format=1.1e1,table-sign-exponent]
S[table-format=3.0]
S[table-format=1.1e1,table-sign-exponent]
S[table-format=1.2]
S[table-format=1.2,table-sign-mantissa]
S[table-format=1.2,table-sign-mantissa]
S[table-format=1.2,table-sign-mantissa]
S[table-format=1.2]
S[table-format=1.3]
S[table-format=1.3]	}\toprule
	state & {$\varepsilon$} & {$\alpha$} & {$\varphi$} & {$\beta$} & {$x_1$} & {$x_2$} & {$x_3$} & {$\delta$} & {$P$} & {$\epsmean$} \\ \hline\\[-8pt]
	$\ket{1,1,1}_{\text{AS}}$ & 2.3e-4 & 1114&1.1e-3 & 0.63 & 1.77 & 0.23 & 2.02& 0.5 & 0.005 & 0.078 \\
	$\ket{\sqrt{2},1.5,2}_{\text{AS}}$ & 4.2e-3 & 341&5.1e-3 & 0.86 & 0.92 & -1.04 & 2.37& 0.4 & 0.002 & 0.082 \\
	$\ket{0.1,4}_{\text B}$ & 1.7e-4 & 481 &2.3e-3 & 0.56 & 0.84 & 1.51 & 2.03 & 0.4 & 0.001 & 0.028 \\
	$\ket{0.4,3}_{\text B}$ & 6e-3 &1105&1.9e-3 & 1.02 & 0 & -1.58 & 1.55& 0.3 & 0.003 & 0.049 \\
	$\ket{0.2,8}_{\text B}$ & 6.3e-3 &449&6.9e-3& 1.54 & 0 & -2.15 & 2.1& 0.5 & 0.004 & 0.054 \\
	$\ket{0.2,10}_{\text B}$ & 1.5e-3 &1679&1.7e-3& 1.42 & 0 & -2.39 & 1.93& 0.8 & 0.002& 0.051 \\
	$\ket{0.3,10}_{\text B}$ & 6.1e-3 & 407 & 8.3e-3 & 1.71 & 0 & -2.43 & 2.44 & 0.7 & 0.001 & 0.057 \\
	$\ket{2,0.1}_{\text{NS}}$ & 5.5e-4 & 671 & 1.3e-3 &  0.45& 0.56 & -2.45 & 0 &  0.15 & 0.009 & 0.088\\
	$\ket{\psi_{02}}$ & 6.4e-4 &227&1.5e-3& 0.17 & -0.09 & -2.99 & 0& 0.3 & 0.003 & 0.062 \\
	$\ket{\psi_{012}}$ & 7.9e-4 &374&5.8e-3& 1.1 & -2.47 & -1.93 & 1.61& 0.5 & 0.008 & 0.087 \\
			\botrule
		\end{tabular*}
\end{table*}

Next, we consider the parameters characterizing the input states, that is, the coherent amplitude $\alpha$ and the phase rotation $\varphi$, which determine the coherent amplitude $\beta$ according to Eq.~\eqref{eq:beta}. Recall that the parameter $\beta$ determines the distance between the coherent states in the generated coherent-state superpositions in the scheme of Fig.~\ref{fig:layout} and it also influences the structure of the output superposition of the setup presented in Fig.~\ref{fig:layout_5}. In coherent-state superpositions approximating quantum states the value of this parameter is generally around 1 \cite{adam2015} ensuring the necessary quantum interference for the generation. From these properties it can be anticipated that the scale of the parameters $\alpha$ and $\varphi$ can be chosen relatively freely applying the only restriction that the corresponding $\beta$ remains around 1.

\begin{table*}[tb]
\caption{Results of the optimization for different nonclassical states for the scheme of Fig.~\ref{fig:layout_5}. The states are approximated by coherent-state superpositions along a line described by Eq.~\eqref{eq:scheme2out}. The table presents for each state the misfit $\varepsilon$ of the approximation and the optimized parameters leading to this misfit including the coherent amplitude $\alpha$, the phase distance $\varphi$, and the resulting coherent amplitude $\beta$, and the squeezing parameter $r$ of the input states, the measurement results $x_i$ of the homodyne measurements, and the overall probability $P$, the corresponding ranges $\delta_i$ of the measurements, and the average misfit $\epsmean$.\label{tab:gabor4}}
\begin{tabular*}{\textwidth}{
l@{\extracolsep{\fill}}
S[table-format=1.1e1,table-sign-exponent]
S[table-format=3.0]
S[table-format=1.1e1,table-sign-exponent]
S[table-format=1.2]
S[table-format=1.3]	
S[table-format=1.2]	
S[table-format=1.2]
S[table-format=1.2]
S[table-format=1.2]
S[table-format=1.3]
S[table-format=1.3]	}\toprule
state & {$\varepsilon$} & {$\alpha$} & {$\varphi$} & {$\beta$} & {$r$} & {$\gamma$} & {$x_1$} & {$x_2$} & {$\delta$} & {$P$} & {$\epsmean$} \\\hline\\[-8pt]
$\ket{1,1,1}_{\text{AS}}$ & 1.7e-3 & 549 & 1.4e-3 & 0.38 & 0.1 & 0.44 & 0.68 & 1.25 & 0.25 & 0.032 & 0.045 \\
$\ket{1,2,1}_{\text{AS}}$ & 3.4e-3 & 1355 & 1.2e-3 & 0.78 & 0.1 & 0.44 & 0.42 & 1.35 & 0.3 & 0.094 & 0.011 \\ 
$|\sqrt{2},3,2\rangle_{\text{AS}}$ & 5.9e-3 & 383 & 6.8e-3 & 1.31 & 0.002 & 0.32 & 0.5 & 0.92 & 0.3 & 0.04 & 0.011 \\
$|0.2,8\rangle_{\text{B}}$ & 6e-3 & 1883 & 1.2e-3 & 1.14 & 0.1 & 0.44 & 0.41 & 1.08 & 0.35 & 0.098 & 0.012 \\
			$|1,0.05\rangle_{\text{NS}}$ & 2.1e-4 & 258 & 2.2e-3 & 0.29 & 0.17 & 0.51 & 1.49 & 2.11 & 0.15 & 0.002 & 0.043 \\
			$|1,0.15\rangle_{\text{NS}}$ & 2.6e-4 & 206 & 6.1e-3 & 0.63 & 0.3 & 0.66 & 1.51 & 2.14 & 0.15 & 0.005 & 0.038 \\ 
			$|2,0.3\rangle_{\text{NS}}$ & 1.3e-5 & 92 & 7.7e-3 & 0.36 & 0.85 & 1.19 & 0.67 & 0 & 0.15 & 0.01 & 0.048 \\ 
			$|2,0.5\rangle_{\text{NS}}$ & 2.8e-3 & 509 & 4.5e-3 & 1.13 & 0.53 & 0.89 & 0.84 & 0 & 0.15 & 0.002& 0.039 \\ 
			$\ket{\psi_{02}}'$ & 9.1e-4 & 731 & 1.1e-3 & 0.38 & 0.002 & 0.32 & 0 & 0 & 0.15 & 0.028 & 0.025 \\
		$\ket{\psi_{012}}''$ & 4.2e-3 & 242 & 3.9e-3 & 0.47 & 0.001 & 0.32 & 0.23 & 1.2 & 0.4 & 0.17 & 0.019 \\\botrule
		\end{tabular*}
\end{table*}

\begin{table*}[tb]
\caption{Results of the optimization for different nonclassical states for the scheme of Fig.~\ref{fig:layout_5}. The states are approximated by coherent-state superpositions on a lattice described by Eq.~\eqref{eq:scheme2out}. The table presents for each state the misfit $\varepsilon$ of the approximation and the optimized parameters leading to this misfit including the coherent amplitude $\alpha$, the phase distance $\varphi$, and the resulting coherent amplitude $\beta$, and the squeezing parameter $r$ of the input states, the measurement results $x_i$ of the homodyne measurements, and the overall probability $P$, the corresponding ranges $\delta_i$ of the measurements, and the average misfit $\epsmean$.\label{tab:gabor5}}
		\begin{tabular*}{\textwidth}{
l@{\extracolsep{\fill}}
S[table-format=1.1e1,table-sign-exponent]
S[table-format=3.0]
S[table-format=1.1e1,table-sign-exponent]
S[table-format=1.2]
S[table-format=1.2]
S[table-format=1.2]	
S[table-format=1.2]
S[table-format=1.2]
S[table-format=1.2]
S[table-format=1.3]
S[table-format=1.3]}\toprule
state &{$\varepsilon$} & {$\alpha$} & {$\varphi$} & {$\beta$} & {$r$} & {$\gamma$} & {$x_1$} & {$x_2$} & {$\delta$} & {$P$} & {$\epsmean$} \\\hline\\[-8pt]
$|1,1,1\rangle_{\text{AS}}$ & 3.2e-4 & 264 & 2.5e-3 & 0.33 & 0.1 & 0.44 & 1.06 & 1.75 & 0.3 & 0.008 & 0.05 \\
$|1,2,1\rangle_{\text{AS}}$ & 3.2e-6 & 586 & 2.7e-3 & 0.78 & 0.13 & 0.47 & 0.42 & 1.5 & 0.35 & 0.13 & 0.003 \\
$|\sqrt{2},1.5,2\rangle_{\text{AS}}$ & 2.5e-3 & 1595 & 1.4e-3 & 1.14 & 0.37 & 0.73 & 1.51 & 1.72 & 0.25 & 0.026 & 0.04 \\
$|0.2,8\rangle_{\text{B}}$ & 6.9e-6 & 1810 & 1.5e-3 & 1.31 & 0.2 & 0.55 & 0 & 1.46 & 0.35 & 0.13 & 0.001\\
$|0.4,6\rangle_{\text{B}}$ & 3.7e-4 & 771 & 4.6e-3 & 1.76 & 0.43 & 0.79 & 1.12 & 1.62 & 0.4 & 0.028 & 0.009\\
$|0.5,5\rangle_{\text{B}}$ & 3.3e-3 & 2751 & 1.3e-3 & 1.83 & 0.54 & 0.9 & 1.49 & 1.65 & 0.45 & 0.014 & 0.028 \\
$\ket{\psi_{02}}'$ & 1.5e-6 & 267 & 4.7e-3 & 0.63 & 0.11 & 0.45 & 0 & 0 & 0.15 & 0.016 & 0.03 \\
$\ket{\psi_{012}}'$ & 9.2e-5 & 90 & 7.5e-3 & 0.33 & 0.1 & 0.44 & 0.11 & 1.04 & 0.2 & 0.051 & 0.003\\ 
$\ket{\psi_{012}}''$ & 1.5e-3 & 403 & 2.7e-3 & 0.55 & 0.13 & 0.47 & 0 & 0.4 & 0.15 & 0.031 & 0.011 \\ \botrule
		\end{tabular*}
\end{table*}

In order to demonstrate this freedom we chose different ranges for $\varphi$ in the scale between $10^{-6}$ to $10^{-1}$ and the coherent amplitudes $\alpha$ in the corresponding ranges determined as described above in the problem of finding optimal parameters for generating different quantum states using our schemes. We found that it is possible to obtain solutions with similar small misfit values for any proper ranges of $\alpha$ and $\varphi$ and for any considered state. An example for this is shown in Tab.~\ref{tab:racs2_1}. In this table we present the results of the optimization for the amplitude squeezed state $\ket{1,2,1}_{\text{AS}}$ approximated by the superposition of Eq.~\eqref{eq:dcss} in the scheme of Fig.~\ref{fig:layout_5}. From the first three rows of the table it can be seen that misfits of the same scale can be achieved for different scales of the parameters $\alpha$ and $\varphi$. Interestingly, solutions of similar misfits can also be obtained even if the optimization is accomplished for different subranges of the same scale of these parameters (see e.g. second three rows of the table). Moreover, this property holds for the last three rows where the optimization is performed only for $\alpha$ and the measurement parameters while fixing the value of $\varphi$ ad hoc. From the data of Tab.~\ref{tab:racs2_1} one can conclude that there exist plenty of pairs of the parameters $\alpha$ and $\varphi$ that can lead to misfits of the same scale. The optimization problem appears to have a large amount of local minimums and it appears to be highly degenerate.
Exploiting this feature of the proposed setups and taking into account the present progress in the topic of cross-Kerr nonlinearities \cite{MatsudaNP2009, HePRA2011, HoiPRL2013, BartkowiakJPB2014}, in the following we choose the order of magnitude of the phase shift $\varphi$ around $10^{-3}$ in our calculations.

In the following tables we present how efficient these schemes are for generating nonclassical states described in the beginning of this Section.
In Tabs.~\ref{tab:emese2} and \ref{tab:emese3} the results of the optimization are shown for the scheme of Fig.~\ref{fig:layout} for approximations by coherent-state superpositions along a line and on a lattice, respectively, described by Eqs.~\eqref{eq:dcss_line} and \eqref{eq:dcss}.
The same is given in Tabs.~\ref{tab:gabor4} and \ref{tab:gabor5} for the scheme of Fig.~\ref{fig:layout_5} for superpositions along a line and on a lattice described by Eq.~\eqref{eq:scheme2out}.
These tables show the misfit of the approximation, all the necessary optimized parameters leading to this misfit including the parameters of the input states, that is, the coherent amplitude $\alpha$, the phase distance $\varphi$, and the squeezing parameter $r$ in the second scheme, the measurement results $x_i$ of the homodyne measurements, the overall probability $P$, the corresponding ranges $\delta_i$ of the measurements, and the average misfit $\epsmean$ for different nonclassical states.

The tabulated data clearly demonstrate that a wide variety of states can be approximated with high precision using our proposed schemes.
Note that there are states which appear in multiple tables, that is, they can be approximated using any of the considered schemes, using various superpositions, albeit with different precision. This is in accordance with the general property of approximations via discrete coherent-state superpositions that they are not unique due to the overcompleteness of coherent states, as mentioned in the Introduction.
The minimal values of the misfits characterizing the accuracy of the approximation that can be achieved vary in the range between $10^{-6}$ and $10^{-3}$.
This precision can be considered to be rather high compared to those that can probably be achieved for the studied states by the quantum state engineering methods based on photon number expansion \cite{fiurasek2005, LeePRA2010}.
It can be partly explained by the high fidelity of the approximation via discrete coherent-state superposition itself \cite{szabo1996, adam2015} on which the proposed schemes are based.

Although the tables show only certain optimal sets of measurement parameters, we note that there are various sets of these parameters leading to the same results for approximating a given state. This degeneracy is implied by the symmetries of the formulas in Eqs.~\eqref{eq:abcoeffs}-\eqref{eq:racs_coeffs}. In the scheme of Fig.~\ref{fig:layout} the same value of the misfit parameter can be achieved by swapping the values of $x_1$ and $x_2$ while changing the sign of $x_3$.
Changing the sign of both $x_1$ and $x_2$ simultaneously does change the value of the misfit neither. We have also noticed that changing only the sign of either $x_1$ or $x_2$ results in the misfit within numerical precision, even though this does not follow from the form of the respective Equations. The same holds for the scheme of Fig.~\ref{fig:layout_5} for the change of the sign of $x_1$.
We have confirmed this latter property for all the approximated states.

The overall probability values presented in the tables were calculated by taking into account the degeneracies of the optimal measurements we have just described. The typical values of this probability is in the range of $10^{-3}-10^{-2}$. We chose the same parameter $\delta$ (the parameter which determines the range of the measurements) for all of the measurements. This parameter was set to a value for which the average misfit was around $10^{-2}$. Obviously, increasing the parameter $\delta$ also increases the overall probability $P$ and the average misfit $\epsmean$ which means that the accuracy of the generation decreases. In Fig.~\ref{fig:eavgP} we present the average misfit $\epsmean$ as a function of the overall probability $P$ for the binomial state $\ket{0.2,10}_{\rm B}$ approximated on a lattice using the scheme presented in Fig.~\ref{fig:layout}. If this relationship is known for a given state, one can find the optimal range $\delta$ of the measurements by deciding upon the optimal balance between the two relevant characteristics of the efficiency: the average misfit and the overall probability. The relatively low values of the measurement probabilities can be explained by the peculiarity of these schemes that they are applicable for the preparation of several states depending on the measurement results of the homodyne measurements. However, these probabilities can be still higher than those achievable in quantum state engineering schemes based on photon addition or subtraction, as they require a larger number of measurements when applied to most of the here considered states~\cite{dakna1999, fiurasek2005}.

\section{Conclusions}\label{sec:concl}

We have proposed two quantum state engineering schemes containing only a few beam splitters and 2 or 3 homodyne measurements for the preparation of nonclassical states based on coherent-state superpositions in traveling optical fields. In spite of their simplicity we find that the schemes are capable of generating a large variety of nonclassical states including amplitude squeezed states, squeezed number states, binomial states and various photon number superpositions.  We demonstrate this by calculating the parameters of the setups to achieve the maximal fidelity of the generated state with respect to the desired one for a large variety of states, and we find it to be relatively high, while the parameters required to achieve them are experimentally feasible. Moreover, the same figures of merit can be achieved with several different choices of parameters of the input states which introduces a freedom to choose a parameter set which is most in line with the characteristics of the applied experimental technology. Meanwhile the success probability is also found to be acceptable.

When compared with photon addition or subtraction based quantum-state engineering schemes, our proposals potentially outperform these in some situations, especially when the states can be expressed as a superposition of a large amount of photon number states. The required number of elements and measurements increases with the number of these states in those schemes, which decreases their success probability and fidelity. In our schemes the required number of elements and measurements is fixed and small.

For traveling optical fields the opportunities introduced by the application of coherent-state superpositions is largely unexplored in spite of the experimental feasibility of the required ingredients in this experimental context. We have demonstrated that our setups of this kind possess features which make them a good candidate for quantum state engineering devices even in practical applications.

\acknowledgements{We thank the support of the National Research, Development and Innovation Office NKFIH (Contract Nos. K124351 and NN109651). This paper is dedicated to the 650th anniversary of the foundation of the University of P\'ecs, Hungary.}

%

%\bibliography{nonclassical_states}

\begin{thebibliography}{55}%
\makeatletter
\providecommand \@ifxundefined [1]{%
 \@ifx{#1\undefined}
}%
\providecommand \@ifnum [1]{%
 \ifnum #1\expandafter \@firstoftwo
 \else \expandafter \@secondoftwo
 \fi
}%
\providecommand \@ifx [1]{%
 \ifx #1\expandafter \@firstoftwo
 \else \expandafter \@secondoftwo
 \fi
}%
\providecommand \natexlab [1]{#1}%
\providecommand \enquote  [1]{``#1''}%
\providecommand \bibnamefont  [1]{#1}%
\providecommand \bibfnamefont [1]{#1}%
\providecommand \citenamefont [1]{#1}%
\providecommand \href@noop [0]{\@secondoftwo}%
\providecommand \href [0]{\begingroup \@sanitize@url \@href}%
\providecommand \@href[1]{\@@startlink{#1}\@@href}%
\providecommand \@@href[1]{\endgroup#1\@@endlink}%
\providecommand \@sanitize@url [0]{\catcode `\\12\catcode `\$12\catcode
  `\&12\catcode `\#12\catcode `\^12\catcode `\_12\catcode `\%12\relax}%
\providecommand \@@startlink[1]{}%
\providecommand \@@endlink[0]{}%
\providecommand \url  [0]{\begingroup\@sanitize@url \@url }%
\providecommand \@url [1]{\endgroup\@href {#1}{\urlprefix }}%
\providecommand \urlprefix  [0]{URL }%
\providecommand \Eprint [0]{\href }%
\providecommand \doibase [0]{http://dx.doi.org/}%
\providecommand \selectlanguage [0]{\@gobble}%
\providecommand \bibinfo  [0]{\@secondoftwo}%
\providecommand \bibfield  [0]{\@secondoftwo}%
\providecommand \translation [1]{[#1]}%
\providecommand \BibitemOpen [0]{}%
\providecommand \bibitemStop [0]{}%
\providecommand \bibitemNoStop [0]{.\EOS\space}%
\providecommand \EOS [0]{\spacefactor3000\relax}%
\providecommand \BibitemShut  [1]{\csname bibitem#1\endcsname}%
\let\auto@bib@innerbib\@empty
%</preamble>
\bibitem [{\citenamefont {Jeong}\ \emph {et~al.}(2004)\citenamefont {Jeong},
  \citenamefont {Kim}, \citenamefont {Ralph},\ and\ \citenamefont
  {Ham}}]{jeong2004}%
  \BibitemOpen
  \bibfield  {author} {\bibinfo {author} {\bibfnamefont {H.}~\bibnamefont
  {Jeong}}, \bibinfo {author} {\bibfnamefont {M.~S.}\ \bibnamefont {Kim}},
  \bibinfo {author} {\bibfnamefont {T.~C.}\ \bibnamefont {Ralph}}, \ and\
  \bibinfo {author} {\bibfnamefont {B.~S.}\ \bibnamefont {Ham}},\ }\href
  {\doibase 10.1103/PhysRevA.70.061801} {\bibfield  {journal} {\bibinfo
  {journal} {Phys. Rev. A}\ }\textbf {\bibinfo {volume} {70}},\ \bibinfo
  {pages} {061801} (\bibinfo {year} {2004})}\BibitemShut {NoStop}%
\bibitem [{\citenamefont {Lund}\ \emph {et~al.}(2004)\citenamefont {Lund},
  \citenamefont {Jeong}, \citenamefont {Ralph},\ and\ \citenamefont
  {Kim}}]{lund2004}%
  \BibitemOpen
  \bibfield  {author} {\bibinfo {author} {\bibfnamefont {A.~P.}\ \bibnamefont
  {Lund}}, \bibinfo {author} {\bibfnamefont {H.}~\bibnamefont {Jeong}},
  \bibinfo {author} {\bibfnamefont {T.~C.}\ \bibnamefont {Ralph}}, \ and\
  \bibinfo {author} {\bibfnamefont {M.~S.}\ \bibnamefont {Kim}},\ }\href@noop
  {} {\bibfield  {journal} {\bibinfo  {journal} {Phys. Rev. A}\ }\textbf
  {\bibinfo {volume} {70}},\ \bibinfo {pages} {020101} (\bibinfo {year}
  {2004})}\BibitemShut {NoStop}%
\bibitem [{\citenamefont {Jeong}\ \emph {et~al.}(2006)\citenamefont {Jeong},
  \citenamefont {Lance}, \citenamefont {Grosse}, \citenamefont {Symul},
  \citenamefont {Lam},\ and\ \citenamefont {Ralph}}]{JeongPRA2006}%
  \BibitemOpen
  \bibfield  {author} {\bibinfo {author} {\bibfnamefont {H.}~\bibnamefont
  {Jeong}}, \bibinfo {author} {\bibfnamefont {A.~M.}\ \bibnamefont {Lance}},
  \bibinfo {author} {\bibfnamefont {N.~B.}\ \bibnamefont {Grosse}}, \bibinfo
  {author} {\bibfnamefont {T.}~\bibnamefont {Symul}}, \bibinfo {author}
  {\bibfnamefont {P.~K.}\ \bibnamefont {Lam}}, \ and\ \bibinfo {author}
  {\bibfnamefont {T.~C.}\ \bibnamefont {Ralph}},\ }\href {\doibase
  10.1103/PhysRevA.74.033813} {\bibfield  {journal} {\bibinfo  {journal} {Phys.
  Rev. A}\ }\textbf {\bibinfo {volume} {74}},\ \bibinfo {pages} {033813}
  (\bibinfo {year} {2006})}\BibitemShut {NoStop}%
\bibitem [{\citenamefont {Lance}\ \emph {et~al.}(2006)\citenamefont {Lance},
  \citenamefont {Jeong}, \citenamefont {Grosse}, \citenamefont {Symul},
  \citenamefont {Ralph},\ and\ \citenamefont {Lam}}]{LancePRA2006}%
  \BibitemOpen
  \bibfield  {author} {\bibinfo {author} {\bibfnamefont {A.~M.}\ \bibnamefont
  {Lance}}, \bibinfo {author} {\bibfnamefont {H.}~\bibnamefont {Jeong}},
  \bibinfo {author} {\bibfnamefont {N.~B.}\ \bibnamefont {Grosse}}, \bibinfo
  {author} {\bibfnamefont {T.}~\bibnamefont {Symul}}, \bibinfo {author}
  {\bibfnamefont {T.~C.}\ \bibnamefont {Ralph}}, \ and\ \bibinfo {author}
  {\bibfnamefont {P.~K.}\ \bibnamefont {Lam}},\ }\href {\doibase
  10.1103/PhysRevA.73.041801} {\bibfield  {journal} {\bibinfo  {journal} {Phys.
  Rev. A}\ }\textbf {\bibinfo {volume} {73}},\ \bibinfo {pages} {041801}
  (\bibinfo {year} {2006})}\BibitemShut {NoStop}%
\bibitem [{\citenamefont {Nielsen}\ and\ \citenamefont
  {M\o{}lmer}(2007)}]{NielsenPRA2007}%
  \BibitemOpen
  \bibfield  {author} {\bibinfo {author} {\bibfnamefont {A.~E.~B.}\
  \bibnamefont {Nielsen}}\ and\ \bibinfo {author} {\bibfnamefont
  {K.}~\bibnamefont {M\o{}lmer}},\ }\href {\doibase 10.1103/PhysRevA.76.043840}
  {\bibfield  {journal} {\bibinfo  {journal} {Phys. Rev. A}\ }\textbf {\bibinfo
  {volume} {76}},\ \bibinfo {pages} {043840} (\bibinfo {year}
  {2007})}\BibitemShut {NoStop}%
\bibitem [{\citenamefont {Glancy}\ and\ \citenamefont
  {Vasconcelos}(2008)}]{glancy2008}%
  \BibitemOpen
  \bibfield  {author} {\bibinfo {author} {\bibfnamefont {S.}~\bibnamefont
  {Glancy}}\ and\ \bibinfo {author} {\bibfnamefont {H.~M.}\ \bibnamefont
  {Vasconcelos}},\ }\href@noop {} {\bibfield  {journal} {\bibinfo  {journal}
  {J. Opt. Soc. Am. B}\ }\textbf {\bibinfo {volume} {25}},\ \bibinfo {pages}
  {712} (\bibinfo {year} {2008})}\BibitemShut {NoStop}%
\bibitem [{\citenamefont {Marek}\ \emph {et~al.}(2008)\citenamefont {Marek},
  \citenamefont {Jeong},\ and\ \citenamefont {Kim}}]{MarekPRA2008}%
  \BibitemOpen
  \bibfield  {author} {\bibinfo {author} {\bibfnamefont {P.}~\bibnamefont
  {Marek}}, \bibinfo {author} {\bibfnamefont {H.}~\bibnamefont {Jeong}}, \ and\
  \bibinfo {author} {\bibfnamefont {M.~S.}\ \bibnamefont {Kim}},\ }\href
  {\doibase 10.1103/PhysRevA.78.063811} {\bibfield  {journal} {\bibinfo
  {journal} {Phys. Rev. A}\ }\textbf {\bibinfo {volume} {78}},\ \bibinfo
  {pages} {063811} (\bibinfo {year} {2008})}\BibitemShut {NoStop}%
\bibitem [{\citenamefont {Takeoka}\ \emph {et~al.}(2008)\citenamefont
  {Takeoka}, \citenamefont {Takahashi},\ and\ \citenamefont
  {Sasaki}}]{TakeokaPRA2008}%
  \BibitemOpen
  \bibfield  {author} {\bibinfo {author} {\bibfnamefont {M.}~\bibnamefont
  {Takeoka}}, \bibinfo {author} {\bibfnamefont {H.}~\bibnamefont {Takahashi}},
  \ and\ \bibinfo {author} {\bibfnamefont {M.}~\bibnamefont {Sasaki}},\ }\href
  {\doibase 10.1103/PhysRevA.77.062315} {\bibfield  {journal} {\bibinfo
  {journal} {Phys. Rev. A}\ }\textbf {\bibinfo {volume} {77}},\ \bibinfo
  {pages} {062315} (\bibinfo {year} {2008})}\BibitemShut {NoStop}%
\bibitem [{\citenamefont {He}\ \emph {et~al.}(2009)\citenamefont {He},
  \citenamefont {Nadeem},\ and\ \citenamefont {Bergou}}]{he2009}%
  \BibitemOpen
  \bibfield  {author} {\bibinfo {author} {\bibfnamefont {B.}~\bibnamefont
  {He}}, \bibinfo {author} {\bibfnamefont {M.}~\bibnamefont {Nadeem}}, \ and\
  \bibinfo {author} {\bibfnamefont {J.}~\bibnamefont {Bergou}},\ }\href@noop {}
  {\bibfield  {journal} {\bibinfo  {journal} {Phys. Rev. A}\ }\textbf {\bibinfo
  {volume} {79}},\ \bibinfo {pages} {035802} (\bibinfo {year}
  {2009})}\BibitemShut {NoStop}%
\bibitem [{\citenamefont {Lee}\ \emph {et~al.}(2012)\citenamefont {Lee},
  \citenamefont {Lee}, \citenamefont {Nha},\ and\ \citenamefont
  {Jeong}}]{LeePRA2012}%
  \BibitemOpen
  \bibfield  {author} {\bibinfo {author} {\bibfnamefont {C.-W.}\ \bibnamefont
  {Lee}}, \bibinfo {author} {\bibfnamefont {J.}~\bibnamefont {Lee}}, \bibinfo
  {author} {\bibfnamefont {H.}~\bibnamefont {Nha}}, \ and\ \bibinfo {author}
  {\bibfnamefont {H.}~\bibnamefont {Jeong}},\ }\href {\doibase
  10.1103/PhysRevA.85.063815} {\bibfield  {journal} {\bibinfo  {journal} {Phys.
  Rev. A}\ }\textbf {\bibinfo {volume} {85}},\ \bibinfo {pages} {063815}
  (\bibinfo {year} {2012})}\BibitemShut {NoStop}%
\bibitem [{\citenamefont {Laghaout}\ \emph {et~al.}(2013)\citenamefont
  {Laghaout}, \citenamefont {Neergaard-Nielsen}, \citenamefont {Rigas},
  \citenamefont {Kragh}, \citenamefont {Tipsmark},\ and\ \citenamefont
  {Andersen}}]{LaghaoutPRA2013}%
  \BibitemOpen
  \bibfield  {author} {\bibinfo {author} {\bibfnamefont {A.}~\bibnamefont
  {Laghaout}}, \bibinfo {author} {\bibfnamefont {J.~S.}\ \bibnamefont
  {Neergaard-Nielsen}}, \bibinfo {author} {\bibfnamefont {I.}~\bibnamefont
  {Rigas}}, \bibinfo {author} {\bibfnamefont {C.}~\bibnamefont {Kragh}},
  \bibinfo {author} {\bibfnamefont {A.}~\bibnamefont {Tipsmark}}, \ and\
  \bibinfo {author} {\bibfnamefont {U.~L.}\ \bibnamefont {Andersen}},\ }\href
  {\doibase 10.1103/PhysRevA.87.043826} {\bibfield  {journal} {\bibinfo
  {journal} {Phys. Rev. A}\ }\textbf {\bibinfo {volume} {87}},\ \bibinfo
  {pages} {043826} (\bibinfo {year} {2013})}\BibitemShut {NoStop}%
\bibitem [{\citenamefont {Wang}\ \emph {et~al.}(2013)\citenamefont {Wang},
  \citenamefont {Yuan},\ and\ \citenamefont {Xu}}]{wang2013}%
  \BibitemOpen
  \bibfield  {author} {\bibinfo {author} {\bibfnamefont {S.}~\bibnamefont
  {Wang}}, \bibinfo {author} {\bibfnamefont {H.-c.}\ \bibnamefont {Yuan}}, \
  and\ \bibinfo {author} {\bibfnamefont {X.-f.}\ \bibnamefont {Xu}},\ }\href
  {\doibase 10.1140/epjd/e2013-30607-7} {\bibfield  {journal} {\bibinfo
  {journal} {Eur. Phys. J. D}\ }\textbf {\bibinfo {volume} {67}},\ \bibinfo
  {eid} {102} (\bibinfo {year} {2013})}\BibitemShut {NoStop}%
\bibitem [{\citenamefont {Jeong}\ \emph {et~al.}(2001)\citenamefont {Jeong},
  \citenamefont {Kim},\ and\ \citenamefont {Lee}}]{jeong2001}%
  \BibitemOpen
  \bibfield  {author} {\bibinfo {author} {\bibfnamefont {H.}~\bibnamefont
  {Jeong}}, \bibinfo {author} {\bibfnamefont {M.~S.}\ \bibnamefont {Kim}}, \
  and\ \bibinfo {author} {\bibfnamefont {J.}~\bibnamefont {Lee}},\ }\href
  {\doibase 10.1103/PhysRevA.64.052308} {\bibfield  {journal} {\bibinfo
  {journal} {Phys. Rev. A}\ }\textbf {\bibinfo {volume} {64}},\ \bibinfo
  {pages} {052308} (\bibinfo {year} {2001})}\BibitemShut {NoStop}%
\bibitem [{\citenamefont {van Enk}\ and\ \citenamefont
  {Hirota}(2001)}]{vanenk2001}%
  \BibitemOpen
  \bibfield  {author} {\bibinfo {author} {\bibfnamefont {S.~J.}\ \bibnamefont
  {van Enk}}\ and\ \bibinfo {author} {\bibfnamefont {O.}~\bibnamefont
  {Hirota}},\ }\href {\doibase 10.1103/PhysRevA.64.022313} {\bibfield
  {journal} {\bibinfo  {journal} {Phys. Rev. A}\ }\textbf {\bibinfo {volume}
  {64}},\ \bibinfo {pages} {022313} (\bibinfo {year} {2001})}\BibitemShut
  {NoStop}%
\bibitem [{\citenamefont {Jeong}\ \emph {et~al.}(2003)\citenamefont {Jeong},
  \citenamefont {Son}, \citenamefont {Kim}, \citenamefont {Ahn},\ and\
  \citenamefont {Brukner}}]{jeong2003}%
  \BibitemOpen
  \bibfield  {author} {\bibinfo {author} {\bibfnamefont {H.}~\bibnamefont
  {Jeong}}, \bibinfo {author} {\bibfnamefont {W.}~\bibnamefont {Son}}, \bibinfo
  {author} {\bibfnamefont {M.~S.}\ \bibnamefont {Kim}}, \bibinfo {author}
  {\bibfnamefont {D.}~\bibnamefont {Ahn}}, \ and\ \bibinfo {author}
  {\bibfnamefont {i.~c.~v.}\ \bibnamefont {Brukner}},\ }\href {\doibase
  10.1103/PhysRevA.67.012106} {\bibfield  {journal} {\bibinfo  {journal} {Phys.
  Rev. A}\ }\textbf {\bibinfo {volume} {67}},\ \bibinfo {pages} {012106}
  (\bibinfo {year} {2003})}\BibitemShut {NoStop}%
\bibitem [{\citenamefont {Ralph}\ \emph {et~al.}(2003)\citenamefont {Ralph},
  \citenamefont {Gilchrist}, \citenamefont {Milburn}, \citenamefont {Munro},\
  and\ \citenamefont {Glancy}}]{ralph2003}%
  \BibitemOpen
  \bibfield  {author} {\bibinfo {author} {\bibfnamefont {T.~C.}\ \bibnamefont
  {Ralph}}, \bibinfo {author} {\bibfnamefont {A.}~\bibnamefont {Gilchrist}},
  \bibinfo {author} {\bibfnamefont {G.~J.}\ \bibnamefont {Milburn}}, \bibinfo
  {author} {\bibfnamefont {W.~J.}\ \bibnamefont {Munro}}, \ and\ \bibinfo
  {author} {\bibfnamefont {S.}~\bibnamefont {Glancy}},\ }\href {\doibase
  10.1103/PhysRevA.68.042319} {\bibfield  {journal} {\bibinfo  {journal} {Phys.
  Rev. A}\ }\textbf {\bibinfo {volume} {68}},\ \bibinfo {pages} {042319}
  (\bibinfo {year} {2003})}\BibitemShut {NoStop}%
\bibitem [{\citenamefont {Asb{\'o}th}\ \emph {et~al.}(2004)\citenamefont
  {Asb{\'o}th}, \citenamefont {Adam}, \citenamefont {Koniorczyk},\ and\
  \citenamefont {Janszky}}]{AsbothEPJD2004}%
  \BibitemOpen
  \bibfield  {author} {\bibinfo {author} {\bibfnamefont {J.~K.}\ \bibnamefont
  {Asb{\'o}th}}, \bibinfo {author} {\bibfnamefont {P.}~\bibnamefont {Adam}},
  \bibinfo {author} {\bibfnamefont {M.}~\bibnamefont {Koniorczyk}}, \ and\
  \bibinfo {author} {\bibfnamefont {J.}~\bibnamefont {Janszky}},\ }\href
  {\doibase 10.1140/epjd/e2004-00094-2} {\bibfield  {journal} {\bibinfo
  {journal} {Eur. Phys. J. D}\ }\textbf {\bibinfo {volume} {30}},\ \bibinfo
  {pages} {403} (\bibinfo {year} {2004})}\BibitemShut {NoStop}%
\bibitem [{\citenamefont {Lund}\ \emph {et~al.}(2008)\citenamefont {Lund},
  \citenamefont {Ralph},\ and\ \citenamefont {Haselgrove}}]{lund2008}%
  \BibitemOpen
  \bibfield  {author} {\bibinfo {author} {\bibfnamefont {A.~P.}\ \bibnamefont
  {Lund}}, \bibinfo {author} {\bibfnamefont {T.~C.}\ \bibnamefont {Ralph}}, \
  and\ \bibinfo {author} {\bibfnamefont {H.~L.}\ \bibnamefont {Haselgrove}},\
  }\href {\doibase 10.1103/PhysRevLett.100.030503} {\bibfield  {journal}
  {\bibinfo  {journal} {Phys. Rev. Lett.}\ }\textbf {\bibinfo {volume} {100}},\
  \bibinfo {pages} {030503} (\bibinfo {year} {2008})}\BibitemShut {NoStop}%
\bibitem [{\citenamefont {Neergaard-Nielsen}\ \emph {et~al.}(2006)\citenamefont
  {Neergaard-Nielsen}, \citenamefont {Melholt~Nielsen}, \citenamefont
  {Hettich}, \citenamefont {M{\o}lmer},\ and\ \citenamefont
  {Polzik}}]{neegard2006}%
  \BibitemOpen
  \bibfield  {author} {\bibinfo {author} {\bibfnamefont {J.~S.}\ \bibnamefont
  {Neergaard-Nielsen}}, \bibinfo {author} {\bibfnamefont {B.}~\bibnamefont
  {Melholt~Nielsen}}, \bibinfo {author} {\bibfnamefont {C.}~\bibnamefont
  {Hettich}}, \bibinfo {author} {\bibfnamefont {K.}~\bibnamefont {M{\o}lmer}},
  \ and\ \bibinfo {author} {\bibfnamefont {E.~S.}\ \bibnamefont {Polzik}},\
  }\href@noop {} {\bibfield  {journal} {\bibinfo  {journal} {Phys. Rev. Lett.}\
  }\textbf {\bibinfo {volume} {97}},\ \bibinfo {pages} {083604} (\bibinfo
  {year} {2006})}\BibitemShut {NoStop}%
\bibitem [{\citenamefont {Ourjoumtsev}\ \emph {et~al.}(2007)\citenamefont
  {Ourjoumtsev}, \citenamefont {Jeong}, \citenamefont {Tualle-Brouri},\ and\
  \citenamefont {Grangier}}]{ourjoumtsev2007}%
  \BibitemOpen
  \bibfield  {author} {\bibinfo {author} {\bibfnamefont {A.}~\bibnamefont
  {Ourjoumtsev}}, \bibinfo {author} {\bibfnamefont {H.}~\bibnamefont {Jeong}},
  \bibinfo {author} {\bibfnamefont {R.}~\bibnamefont {Tualle-Brouri}}, \ and\
  \bibinfo {author} {\bibfnamefont {P.}~\bibnamefont {Grangier}},\ }\href@noop
  {} {\bibfield  {journal} {\bibinfo  {journal} {Nature}\ }\textbf {\bibinfo
  {volume} {448}} (\bibinfo {year} {2007})}\BibitemShut {NoStop}%
\bibitem [{\citenamefont {Takahashi}\ \emph {et~al.}(2008)\citenamefont
  {Takahashi}, \citenamefont {Wakui}, \citenamefont {Suzuki}, \citenamefont
  {Takeoka}, \citenamefont {Hayasaka}, \citenamefont {Furusawa},\ and\
  \citenamefont {Sasaki}}]{TakahashiPRL2008}%
  \BibitemOpen
  \bibfield  {author} {\bibinfo {author} {\bibfnamefont {H.}~\bibnamefont
  {Takahashi}}, \bibinfo {author} {\bibfnamefont {K.}~\bibnamefont {Wakui}},
  \bibinfo {author} {\bibfnamefont {S.}~\bibnamefont {Suzuki}}, \bibinfo
  {author} {\bibfnamefont {M.}~\bibnamefont {Takeoka}}, \bibinfo {author}
  {\bibfnamefont {K.}~\bibnamefont {Hayasaka}}, \bibinfo {author}
  {\bibfnamefont {A.}~\bibnamefont {Furusawa}}, \ and\ \bibinfo {author}
  {\bibfnamefont {M.}~\bibnamefont {Sasaki}},\ }\href {\doibase
  10.1103/PhysRevLett.101.233605} {\bibfield  {journal} {\bibinfo  {journal}
  {Phys. Rev. Lett.}\ }\textbf {\bibinfo {volume} {101}},\ \bibinfo {pages}
  {233605} (\bibinfo {year} {2008})}\BibitemShut {NoStop}%
\bibitem [{\citenamefont {Gerrits}\ \emph {et~al.}(2010)\citenamefont
  {Gerrits}, \citenamefont {Glancy}, \citenamefont {Clement}, \citenamefont
  {Calkins}, \citenamefont {Lita}, \citenamefont {Miller}, \citenamefont
  {Migdall}, \citenamefont {Nam}, \citenamefont {Mirin},\ and\ \citenamefont
  {Knill}}]{GerritsPRA2010}%
  \BibitemOpen
  \bibfield  {author} {\bibinfo {author} {\bibfnamefont {T.}~\bibnamefont
  {Gerrits}}, \bibinfo {author} {\bibfnamefont {S.}~\bibnamefont {Glancy}},
  \bibinfo {author} {\bibfnamefont {T.~S.}\ \bibnamefont {Clement}}, \bibinfo
  {author} {\bibfnamefont {B.}~\bibnamefont {Calkins}}, \bibinfo {author}
  {\bibfnamefont {A.~E.}\ \bibnamefont {Lita}}, \bibinfo {author}
  {\bibfnamefont {A.~J.}\ \bibnamefont {Miller}}, \bibinfo {author}
  {\bibfnamefont {A.~L.}\ \bibnamefont {Migdall}}, \bibinfo {author}
  {\bibfnamefont {S.~W.}\ \bibnamefont {Nam}}, \bibinfo {author} {\bibfnamefont
  {R.~P.}\ \bibnamefont {Mirin}}, \ and\ \bibinfo {author} {\bibfnamefont
  {E.}~\bibnamefont {Knill}},\ }\href {\doibase 10.1103/PhysRevA.82.031802}
  {\bibfield  {journal} {\bibinfo  {journal} {Phys. Rev. A}\ }\textbf {\bibinfo
  {volume} {82}},\ \bibinfo {pages} {031802} (\bibinfo {year}
  {2010})}\BibitemShut {NoStop}%
\bibitem [{\citenamefont {Huang}\ \emph {et~al.}(2015)\citenamefont {Huang},
  \citenamefont {Le~Jeannic}, \citenamefont {Ruaudel}, \citenamefont {Verma},
  \citenamefont {Shaw}, \citenamefont {Marsili}, \citenamefont {Nam},
  \citenamefont {Wu}, \citenamefont {Zeng}, \citenamefont {Jeong},
  \citenamefont {Filip}, \citenamefont {Morin},\ and\ \citenamefont
  {Laurat}}]{HuangPRL2015}%
  \BibitemOpen
  \bibfield  {author} {\bibinfo {author} {\bibfnamefont {K.}~\bibnamefont
  {Huang}}, \bibinfo {author} {\bibfnamefont {H.}~\bibnamefont {Le~Jeannic}},
  \bibinfo {author} {\bibfnamefont {J.}~\bibnamefont {Ruaudel}}, \bibinfo
  {author} {\bibfnamefont {V.~B.}\ \bibnamefont {Verma}}, \bibinfo {author}
  {\bibfnamefont {M.~D.}\ \bibnamefont {Shaw}}, \bibinfo {author}
  {\bibfnamefont {F.}~\bibnamefont {Marsili}}, \bibinfo {author} {\bibfnamefont
  {S.~W.}\ \bibnamefont {Nam}}, \bibinfo {author} {\bibfnamefont
  {E.}~\bibnamefont {Wu}}, \bibinfo {author} {\bibfnamefont {H.}~\bibnamefont
  {Zeng}}, \bibinfo {author} {\bibfnamefont {Y.-C.}\ \bibnamefont {Jeong}},
  \bibinfo {author} {\bibfnamefont {R.}~\bibnamefont {Filip}}, \bibinfo
  {author} {\bibfnamefont {O.}~\bibnamefont {Morin}}, \ and\ \bibinfo {author}
  {\bibfnamefont {J.}~\bibnamefont {Laurat}},\ }\href {\doibase
  10.1103/PhysRevLett.115.023602} {\bibfield  {journal} {\bibinfo  {journal}
  {Phys. Rev. Lett.}\ }\textbf {\bibinfo {volume} {115}},\ \bibinfo {pages}
  {023602} (\bibinfo {year} {2015})}\BibitemShut {NoStop}%
\bibitem [{\citenamefont {Szabo}\ \emph {et~al.}(1996)\citenamefont {Szabo},
  \citenamefont {Adam}, \citenamefont {Janszky},\ and\ \citenamefont
  {Domokos}}]{szabo1996}%
  \BibitemOpen
  \bibfield  {author} {\bibinfo {author} {\bibfnamefont {S.}~\bibnamefont
  {Szabo}}, \bibinfo {author} {\bibfnamefont {P.}~\bibnamefont {Adam}},
  \bibinfo {author} {\bibfnamefont {J.}~\bibnamefont {Janszky}}, \ and\
  \bibinfo {author} {\bibfnamefont {P.}~\bibnamefont {Domokos}},\ }\href@noop
  {} {\bibfield  {journal} {\bibinfo  {journal} {Phys. Rev. A}\ }\textbf
  {\bibinfo {volume} {53}},\ \bibinfo {pages} {2698} (\bibinfo {year}
  {1996})}\BibitemShut {NoStop}%
\bibitem [{\citenamefont {Dakna}\ \emph {et~al.}(1999)\citenamefont {Dakna},
  \citenamefont {Clausen}, \citenamefont {Kn\"{o}ll},\ and\ \citenamefont
  {Welsch}}]{dakna1999}%
  \BibitemOpen
  \bibfield  {author} {\bibinfo {author} {\bibfnamefont {M.}~\bibnamefont
  {Dakna}}, \bibinfo {author} {\bibfnamefont {J.}~\bibnamefont {Clausen}},
  \bibinfo {author} {\bibfnamefont {L.}~\bibnamefont {Kn\"{o}ll}}, \ and\
  \bibinfo {author} {\bibfnamefont {D.-G.}\ \bibnamefont {Welsch}},\
  }\href@noop {} {\bibfield  {journal} {\bibinfo  {journal} {Phys. Rev. A}\
  }\textbf {\bibinfo {volume} {59}},\ \bibinfo {pages} {1658} (\bibinfo {year}
  {1999})}\BibitemShut {NoStop}%
\bibitem [{\citenamefont {Fiur\'{a}sek}\ \emph {et~al.}(2005)\citenamefont
  {Fiur\'{a}sek}, \citenamefont {G\'{a}rcia-Patr\'{o}n},\ and\ \citenamefont
  {Cerf}}]{fiurasek2005}%
  \BibitemOpen
  \bibfield  {author} {\bibinfo {author} {\bibfnamefont {J.}~\bibnamefont
  {Fiur\'{a}sek}}, \bibinfo {author} {\bibfnamefont {R.}~\bibnamefont
  {G\'{a}rcia-Patr\'{o}n}}, \ and\ \bibinfo {author} {\bibfnamefont {N.~J.}\
  \bibnamefont {Cerf}},\ }\href@noop {} {\bibfield  {journal} {\bibinfo
  {journal} {Phys. Rev. A}\ }\textbf {\bibinfo {volume} {72}},\ \bibinfo
  {pages} {033822} (\bibinfo {year} {2005})}\BibitemShut {NoStop}%
\bibitem [{\citenamefont {Gerry}\ and\ \citenamefont
  {Benmoussa}(2006)}]{GerryPRA2006}%
  \BibitemOpen
  \bibfield  {author} {\bibinfo {author} {\bibfnamefont {C.~C.}\ \bibnamefont
  {Gerry}}\ and\ \bibinfo {author} {\bibfnamefont {A.}~\bibnamefont
  {Benmoussa}},\ }\href {\doibase 10.1103/PhysRevA.73.063817} {\bibfield
  {journal} {\bibinfo  {journal} {Phys. Rev. A}\ }\textbf {\bibinfo {volume}
  {73}},\ \bibinfo {pages} {063817} (\bibinfo {year} {2006})}\BibitemShut
  {NoStop}%
\bibitem [{\citenamefont {Bimbard}\ \emph {et~al.}(2010)\citenamefont
  {Bimbard}, \citenamefont {Jain}, \citenamefont {MacRae},\ and\ \citenamefont
  {Lvovsky}}]{BimbardNP2010}%
  \BibitemOpen
  \bibfield  {author} {\bibinfo {author} {\bibfnamefont {E.}~\bibnamefont
  {Bimbard}}, \bibinfo {author} {\bibfnamefont {N.}~\bibnamefont {Jain}},
  \bibinfo {author} {\bibfnamefont {A.}~\bibnamefont {MacRae}}, \ and\ \bibinfo
  {author} {\bibfnamefont {A.~I.}\ \bibnamefont {Lvovsky}},\ }\href
  {http://dx.doi.org/10.1038/nphoton.2010.6} {\bibfield  {journal} {\bibinfo
  {journal} {Nat. Photon.}\ }\textbf {\bibinfo {volume} {4}},\ \bibinfo {pages}
  {243} (\bibinfo {year} {2010})}\BibitemShut {NoStop}%
\bibitem [{\citenamefont {Lee}\ and\ \citenamefont {Nha}(2010)}]{LeePRA2010}%
  \BibitemOpen
  \bibfield  {author} {\bibinfo {author} {\bibfnamefont {S.-Y.}\ \bibnamefont
  {Lee}}\ and\ \bibinfo {author} {\bibfnamefont {H.}~\bibnamefont {Nha}},\
  }\href {\doibase 10.1103/PhysRevA.82.053812} {\bibfield  {journal} {\bibinfo
  {journal} {Phys. Rev. A}\ }\textbf {\bibinfo {volume} {82}},\ \bibinfo
  {pages} {053812} (\bibinfo {year} {2010})}\BibitemShut {NoStop}%
\bibitem [{\citenamefont {Sperling}\ \emph {et~al.}(2014)\citenamefont
  {Sperling}, \citenamefont {Vogel},\ and\ \citenamefont
  {Agarwal}}]{SperlingPRA2014}%
  \BibitemOpen
  \bibfield  {author} {\bibinfo {author} {\bibfnamefont {J.}~\bibnamefont
  {Sperling}}, \bibinfo {author} {\bibfnamefont {W.}~\bibnamefont {Vogel}}, \
  and\ \bibinfo {author} {\bibfnamefont {G.~S.}\ \bibnamefont {Agarwal}},\
  }\href {\doibase 10.1103/PhysRevA.89.043829} {\bibfield  {journal} {\bibinfo
  {journal} {Phys. Rev. A}\ }\textbf {\bibinfo {volume} {89}},\ \bibinfo
  {pages} {043829} (\bibinfo {year} {2014})}\BibitemShut {NoStop}%
\bibitem [{\citenamefont {Adam}\ \emph {et~al.}(2015)\citenamefont {Adam},
  \citenamefont {Molnar}, \citenamefont {Mogyorosi}, \citenamefont {Varga},
  \citenamefont {Mechler},\ and\ \citenamefont {Janszky}}]{adam2015}%
  \BibitemOpen
  \bibfield  {author} {\bibinfo {author} {\bibfnamefont {P.}~\bibnamefont
  {Adam}}, \bibinfo {author} {\bibfnamefont {E.}~\bibnamefont {Molnar}},
  \bibinfo {author} {\bibfnamefont {G.}~\bibnamefont {Mogyorosi}}, \bibinfo
  {author} {\bibfnamefont {A.}~\bibnamefont {Varga}}, \bibinfo {author}
  {\bibfnamefont {M.}~\bibnamefont {Mechler}}, \ and\ \bibinfo {author}
  {\bibfnamefont {J.}~\bibnamefont {Janszky}},\ }\href@noop {} {\bibfield
  {journal} {\bibinfo  {journal} {Phys. Scr.}\ }\textbf {\bibinfo {volume}
  {90}},\ \bibinfo {pages} {074021} (\bibinfo {year} {2015})}\BibitemShut
  {NoStop}%
\bibitem [{\citenamefont {Huang}\ \emph {et~al.}(2016)\citenamefont {Huang},
  \citenamefont {{Le Jeannic}}, \citenamefont {Verma}, \citenamefont {Shaw},
  \citenamefont {Marsili}, \citenamefont {Nam}, \citenamefont {Wu},
  \citenamefont {Zeng}, \citenamefont {Morin},\ and\ \citenamefont
  {Laurat}}]{HuangPRA2016}%
  \BibitemOpen
  \bibfield  {author} {\bibinfo {author} {\bibfnamefont {K.}~\bibnamefont
  {Huang}}, \bibinfo {author} {\bibfnamefont {H.}~\bibnamefont {{Le Jeannic}}},
  \bibinfo {author} {\bibfnamefont {V.~B.}\ \bibnamefont {Verma}}, \bibinfo
  {author} {\bibfnamefont {M.~D.}\ \bibnamefont {Shaw}}, \bibinfo {author}
  {\bibfnamefont {F.}~\bibnamefont {Marsili}}, \bibinfo {author} {\bibfnamefont
  {S.~W.}\ \bibnamefont {Nam}}, \bibinfo {author} {\bibfnamefont
  {E.}~\bibnamefont {Wu}}, \bibinfo {author} {\bibfnamefont {H.}~\bibnamefont
  {Zeng}}, \bibinfo {author} {\bibfnamefont {O.}~\bibnamefont {Morin}}, \ and\
  \bibinfo {author} {\bibfnamefont {J.}~\bibnamefont {Laurat}},\ }\href
  {\doibase 10.1103/PhysRevA.93.013838} {\bibfield  {journal} {\bibinfo
  {journal} {Phys. Rev. A}\ }\textbf {\bibinfo {volume} {93}},\ \bibinfo
  {pages} {013838} (\bibinfo {year} {2016})}\BibitemShut {NoStop}%
\bibitem [{\citenamefont {Janszky}\ \emph {et~al.}(1993)\citenamefont
  {Janszky}, \citenamefont {Domokos},\ and\ \citenamefont
  {Adam}}]{janszky1993}%
  \BibitemOpen
  \bibfield  {author} {\bibinfo {author} {\bibfnamefont {J.}~\bibnamefont
  {Janszky}}, \bibinfo {author} {\bibfnamefont {P.}~\bibnamefont {Domokos}}, \
  and\ \bibinfo {author} {\bibfnamefont {P.}~\bibnamefont {Adam}},\ }\href@noop
  {} {\bibfield  {journal} {\bibinfo  {journal} {Phys. Rev. A}\ }\textbf
  {\bibinfo {volume} {48}},\ \bibinfo {pages} {2213} (\bibinfo {year}
  {1993})}\BibitemShut {NoStop}%
\bibitem [{\citenamefont {Janszky}\ \emph {et~al.}(1995)\citenamefont
  {Janszky}, \citenamefont {Domokos}, \citenamefont {Szabo},\ and\
  \citenamefont {Adam}}]{janszky1995}%
  \BibitemOpen
  \bibfield  {author} {\bibinfo {author} {\bibfnamefont {J.}~\bibnamefont
  {Janszky}}, \bibinfo {author} {\bibfnamefont {P.}~\bibnamefont {Domokos}},
  \bibinfo {author} {\bibfnamefont {S.}~\bibnamefont {Szabo}}, \ and\ \bibinfo
  {author} {\bibfnamefont {P.}~\bibnamefont {Adam}},\ }\href@noop {} {\bibfield
   {journal} {\bibinfo  {journal} {Phys. Rev. A}\ }\textbf {\bibinfo {volume}
  {51}},\ \bibinfo {pages} {4191} (\bibinfo {year} {1995})}\BibitemShut
  {NoStop}%
\bibitem [{\citenamefont {Stergioulas}\ \emph {et~al.}(1999)\citenamefont
  {Stergioulas}, \citenamefont {Vassiliadis},\ and\ \citenamefont
  {Vourdas}}]{stergioulas1999}%
  \BibitemOpen
  \bibfield  {author} {\bibinfo {author} {\bibfnamefont {L.~K.}\ \bibnamefont
  {Stergioulas}}, \bibinfo {author} {\bibfnamefont {V.~S.}\ \bibnamefont
  {Vassiliadis}}, \ and\ \bibinfo {author} {\bibfnamefont {A.}~\bibnamefont
  {Vourdas}},\ }\href {http://stacks.iop.org/0305-4470/32/i=17/a=308}
  {\bibfield  {journal} {\bibinfo  {journal} {J. Phys. A: Math. Gen.}\ }\textbf
  {\bibinfo {volume} {32}},\ \bibinfo {pages} {3169} (\bibinfo {year}
  {1999})}\BibitemShut {NoStop}%
\bibitem [{\citenamefont {Domokos}\ \emph {et~al.}(1994)\citenamefont
  {Domokos}, \citenamefont {Janszky},\ and\ \citenamefont
  {Adam}}]{DomokosPRA1994}%
  \BibitemOpen
  \bibfield  {author} {\bibinfo {author} {\bibfnamefont {P.}~\bibnamefont
  {Domokos}}, \bibinfo {author} {\bibfnamefont {J.}~\bibnamefont {Janszky}}, \
  and\ \bibinfo {author} {\bibfnamefont {P.}~\bibnamefont {Adam}},\ }\href
  {\doibase 10.1103/PhysRevA.50.3340} {\bibfield  {journal} {\bibinfo
  {journal} {Phys. Rev. A}\ }\textbf {\bibinfo {volume} {50}},\ \bibinfo
  {pages} {3340} (\bibinfo {year} {1994})}\BibitemShut {NoStop}%
\bibitem [{\citenamefont {Lutterbach}\ and\ \citenamefont
  {Davidovich}(2000)}]{LutterbachPRA2000}%
  \BibitemOpen
  \bibfield  {author} {\bibinfo {author} {\bibfnamefont {L.~G.}\ \bibnamefont
  {Lutterbach}}\ and\ \bibinfo {author} {\bibfnamefont {L.}~\bibnamefont
  {Davidovich}},\ }\href {\doibase 10.1103/PhysRevA.61.023813} {\bibfield
  {journal} {\bibinfo  {journal} {Phys. Rev. A}\ }\textbf {\bibinfo {volume}
  {61}},\ \bibinfo {pages} {023813} (\bibinfo {year} {2000})}\BibitemShut
  {NoStop}%
\bibitem [{\citenamefont {Pathak}\ and\ \citenamefont
  {Agarwal}(2005)}]{PathakPRA2005}%
  \BibitemOpen
  \bibfield  {author} {\bibinfo {author} {\bibfnamefont {P.~K.}\ \bibnamefont
  {Pathak}}\ and\ \bibinfo {author} {\bibfnamefont {G.~S.}\ \bibnamefont
  {Agarwal}},\ }\href {\doibase 10.1103/PhysRevA.71.043823} {\bibfield
  {journal} {\bibinfo  {journal} {Phys. Rev. A}\ }\textbf {\bibinfo {volume}
  {71}},\ \bibinfo {pages} {043823} (\bibinfo {year} {2005})}\BibitemShut
  {NoStop}%
\bibitem [{\citenamefont {Moya-Cessa}\ \emph {et~al.}(1999)\citenamefont
  {Moya-Cessa}, \citenamefont {Wallentowitz},\ and\ \citenamefont
  {Vogel}}]{MoyaCessaPRA1999}%
  \BibitemOpen
  \bibfield  {author} {\bibinfo {author} {\bibfnamefont {H.}~\bibnamefont
  {Moya-Cessa}}, \bibinfo {author} {\bibfnamefont {S.}~\bibnamefont
  {Wallentowitz}}, \ and\ \bibinfo {author} {\bibfnamefont {W.}~\bibnamefont
  {Vogel}},\ }\href {\doibase 10.1103/PhysRevA.59.2920} {\bibfield  {journal}
  {\bibinfo  {journal} {Phys. Rev. A}\ }\textbf {\bibinfo {volume} {59}},\
  \bibinfo {pages} {2920} (\bibinfo {year} {1999})}\BibitemShut {NoStop}%
\bibitem [{\citenamefont {Gerry}(1999)}]{gerry1999}%
  \BibitemOpen
  \bibfield  {author} {\bibinfo {author} {\bibfnamefont {C.~C.}\ \bibnamefont
  {Gerry}},\ }\href@noop {} {\bibfield  {journal} {\bibinfo  {journal} {Phys.
  Rev. A}\ }\textbf {\bibinfo {volume} {59}},\ \bibinfo {pages} {4095}
  (\bibinfo {year} {1999})}\BibitemShut {NoStop}%
\bibitem [{\citenamefont {Jeong}(2005)}]{jeong2005}%
  \BibitemOpen
  \bibfield  {author} {\bibinfo {author} {\bibfnamefont {H.}~\bibnamefont
  {Jeong}},\ }\href@noop {} {\bibfield  {journal} {\bibinfo  {journal} {Phys.
  Rev. A}\ }\textbf {\bibinfo {volume} {72}},\ \bibinfo {pages} {034305}
  (\bibinfo {year} {2005})}\BibitemShut {NoStop}%
\bibitem [{\citenamefont {Matsuda}\ \emph {et~al.}(2009)\citenamefont
  {Matsuda}, \citenamefont {Shimizu}, \citenamefont {Mitsumori}, \citenamefont
  {Kosaka},\ and\ \citenamefont {Edamatsu}}]{MatsudaNP2009}%
  \BibitemOpen
  \bibfield  {author} {\bibinfo {author} {\bibfnamefont {N.}~\bibnamefont
  {Matsuda}}, \bibinfo {author} {\bibfnamefont {R.}~\bibnamefont {Shimizu}},
  \bibinfo {author} {\bibfnamefont {Y.}~\bibnamefont {Mitsumori}}, \bibinfo
  {author} {\bibfnamefont {H.}~\bibnamefont {Kosaka}}, \ and\ \bibinfo {author}
  {\bibfnamefont {K.}~\bibnamefont {Edamatsu}},\ }\href
  {http://dx.doi.org/10.1038/nphoton.2008.292} {\bibfield  {journal} {\bibinfo
  {journal} {Nat. Photon.}\ }\textbf {\bibinfo {volume} {3}},\ \bibinfo {pages}
  {95} (\bibinfo {year} {2009})}\BibitemShut {NoStop}%
\bibitem [{\citenamefont {He}\ \emph {et~al.}(2011)\citenamefont {He},
  \citenamefont {Lin},\ and\ \citenamefont {Simon}}]{HePRA2011}%
  \BibitemOpen
  \bibfield  {author} {\bibinfo {author} {\bibfnamefont {B.}~\bibnamefont
  {He}}, \bibinfo {author} {\bibfnamefont {Q.}~\bibnamefont {Lin}}, \ and\
  \bibinfo {author} {\bibfnamefont {C.}~\bibnamefont {Simon}},\ }\href
  {\doibase 10.1103/PhysRevA.83.053826} {\bibfield  {journal} {\bibinfo
  {journal} {Phys. Rev. A}\ }\textbf {\bibinfo {volume} {83}},\ \bibinfo
  {pages} {053826} (\bibinfo {year} {2011})}\BibitemShut {NoStop}%
\bibitem [{\citenamefont {Hoi}\ \emph {et~al.}(2013)\citenamefont {Hoi},
  \citenamefont {Kockum}, \citenamefont {Palomaki}, \citenamefont {Stace},
  \citenamefont {Fan}, \citenamefont {Tornberg}, \citenamefont {Sathyamoorthy},
  \citenamefont {Johansson}, \citenamefont {Delsing},\ and\ \citenamefont
  {Wilson}}]{HoiPRL2013}%
  \BibitemOpen
  \bibfield  {author} {\bibinfo {author} {\bibfnamefont {I.-C.}\ \bibnamefont
  {Hoi}}, \bibinfo {author} {\bibfnamefont {A.~F.}\ \bibnamefont {Kockum}},
  \bibinfo {author} {\bibfnamefont {T.}~\bibnamefont {Palomaki}}, \bibinfo
  {author} {\bibfnamefont {T.~M.}\ \bibnamefont {Stace}}, \bibinfo {author}
  {\bibfnamefont {B.}~\bibnamefont {Fan}}, \bibinfo {author} {\bibfnamefont
  {L.}~\bibnamefont {Tornberg}}, \bibinfo {author} {\bibfnamefont {S.~R.}\
  \bibnamefont {Sathyamoorthy}}, \bibinfo {author} {\bibfnamefont
  {G.}~\bibnamefont {Johansson}}, \bibinfo {author} {\bibfnamefont
  {P.}~\bibnamefont {Delsing}}, \ and\ \bibinfo {author} {\bibfnamefont
  {C.~M.}\ \bibnamefont {Wilson}},\ }\href {\doibase
  10.1103/PhysRevLett.111.053601} {\bibfield  {journal} {\bibinfo  {journal}
  {Phys. Rev. Lett.}\ }\textbf {\bibinfo {volume} {111}},\ \bibinfo {pages}
  {053601} (\bibinfo {year} {2013})}\BibitemShut {NoStop}%
\bibitem [{\citenamefont {Bartkowiak}\ \emph {et~al.}(2014)\citenamefont
  {Bartkowiak}, \citenamefont {Wu},\ and\ \citenamefont
  {Miranowicz}}]{BartkowiakJPB2014}%
  \BibitemOpen
  \bibfield  {author} {\bibinfo {author} {\bibfnamefont {M.}~\bibnamefont
  {Bartkowiak}}, \bibinfo {author} {\bibfnamefont {L.-A.}\ \bibnamefont {Wu}},
  \ and\ \bibinfo {author} {\bibfnamefont {A.}~\bibnamefont {Miranowicz}},\
  }\href {http://stacks.iop.org/0953-4075/47/i=14/a=145501} {\bibfield
  {journal} {\bibinfo  {journal} {J. Phys. B: At., Mol. Opt. Phys.}\ }\textbf
  {\bibinfo {volume} {47}},\ \bibinfo {pages} {145501} (\bibinfo {year}
  {2014})}\BibitemShut {NoStop}%
\bibitem [{\citenamefont {Adam}\ \emph {et~al.}(2010)\citenamefont {Adam},
  \citenamefont {Kiss}, \citenamefont {Dar\'azs},\ and\ \citenamefont
  {Jex}}]{adam2010}%
  \BibitemOpen
  \bibfield  {author} {\bibinfo {author} {\bibfnamefont {P.}~\bibnamefont
  {Adam}}, \bibinfo {author} {\bibfnamefont {T.}~\bibnamefont {Kiss}}, \bibinfo
  {author} {\bibfnamefont {Z.}~\bibnamefont {Dar\'azs}}, \ and\ \bibinfo
  {author} {\bibfnamefont {I.}~\bibnamefont {Jex}},\ }\href@noop {} {\bibfield
  {journal} {\bibinfo  {journal} {Phys. Scr.}\ }\textbf {\bibinfo {volume}
  {T140}},\ \bibinfo {pages} {014011} (\bibinfo {year} {2010})}\BibitemShut
  {NoStop}%
\bibitem [{\citenamefont {Marek}\ and\ \citenamefont
  {Fiur\'a\ifmmode~\check{s}\else \v{s}\fi{}ek}(2009)}]{MarekPRA2009}%
  \BibitemOpen
  \bibfield  {author} {\bibinfo {author} {\bibfnamefont {P.}~\bibnamefont
  {Marek}}\ and\ \bibinfo {author} {\bibfnamefont {J.}~\bibnamefont
  {Fiur\'a\ifmmode~\check{s}\else \v{s}\fi{}ek}},\ }\href {\doibase
  10.1103/PhysRevA.79.062321} {\bibfield  {journal} {\bibinfo  {journal} {Phys.
  Rev. A}\ }\textbf {\bibinfo {volume} {79}},\ \bibinfo {pages} {062321}
  (\bibinfo {year} {2009})}\BibitemShut {NoStop}%
\bibitem [{\citenamefont {Adam}\ \emph {et~al.}(1991)\citenamefont {Adam},
  \citenamefont {Jansky},\ and\ \citenamefont {Vinogradov}}]{AdamPLA1991}%
  \BibitemOpen
  \bibfield  {author} {\bibinfo {author} {\bibfnamefont {P.}~\bibnamefont
  {Adam}}, \bibinfo {author} {\bibfnamefont {J.}~\bibnamefont {Jansky}}, \ and\
  \bibinfo {author} {\bibfnamefont {A.}~\bibnamefont {Vinogradov}},\ }\href
  {\doibase http://dx.doi.org/10.1016/0375-9601(91)91057-K} {\bibfield
  {journal} {\bibinfo  {journal} {Physics Letters A}\ }\textbf {\bibinfo
  {volume} {160}},\ \bibinfo {pages} {506 } (\bibinfo {year}
  {1991})}\BibitemShut {NoStop}%
\bibitem [{\citenamefont {Adam}\ \emph {et~al.}(2014)\citenamefont {Adam},
  \citenamefont {Mechler}, \citenamefont {Szalay},\ and\ \citenamefont
  {Koniorczyk}}]{AdamSzalayPRA2014}%
  \BibitemOpen
  \bibfield  {author} {\bibinfo {author} {\bibfnamefont {P.}~\bibnamefont
  {Adam}}, \bibinfo {author} {\bibfnamefont {M.}~\bibnamefont {Mechler}},
  \bibinfo {author} {\bibfnamefont {V.}~\bibnamefont {Szalay}}, \ and\ \bibinfo
  {author} {\bibfnamefont {M.}~\bibnamefont {Koniorczyk}},\ }\href {\doibase
  10.1103/PhysRevA.89.062108} {\bibfield  {journal} {\bibinfo  {journal} {Phys.
  Rev. A}\ }\textbf {\bibinfo {volume} {89}},\ \bibinfo {pages} {062108}
  (\bibinfo {year} {2014})}\BibitemShut {NoStop}%
\bibitem [{\citenamefont {Stoler}\ \emph {et~al.}(1985)\citenamefont {Stoler},
  \citenamefont {Saleh},\ and\ \citenamefont {Teich}}]{Stoler1985}%
  \BibitemOpen
  \bibfield  {author} {\bibinfo {author} {\bibfnamefont {D.}~\bibnamefont
  {Stoler}}, \bibinfo {author} {\bibfnamefont {B.~E.~A.}\ \bibnamefont
  {Saleh}}, \ and\ \bibinfo {author} {\bibfnamefont {M.~C.}\ \bibnamefont
  {Teich}},\ }\href {\doibase 10.1080/713821735} {\bibfield  {journal}
  {\bibinfo  {journal} {Opt. Acta}\ }\textbf {\bibinfo {volume} {32}},\
  \bibinfo {pages} {345} (\bibinfo {year} {1985})}\BibitemShut {NoStop}%
\bibitem [{\citenamefont {Kim}\ \emph {et~al.}(1989)\citenamefont {Kim},
  \citenamefont {Oliviera},\ and\ \citenamefont {Knight}}]{kim1989}%
  \BibitemOpen
  \bibfield  {author} {\bibinfo {author} {\bibfnamefont {M.~S.}\ \bibnamefont
  {Kim}}, \bibinfo {author} {\bibfnamefont {F.~A.~M.}\ \bibnamefont
  {Oliviera}}, \ and\ \bibinfo {author} {\bibfnamefont {P.~L.}\ \bibnamefont
  {Knight}},\ }\href@noop {} {\bibfield  {journal} {\bibinfo  {journal} {Phys.
  Rev. A}\ }\textbf {\bibinfo {volume} {40}},\ \bibinfo {pages} {2494}
  (\bibinfo {year} {1989})}\BibitemShut {NoStop}%
\bibitem [{\citenamefont {Liu}\ and\ \citenamefont {Tombesi}(1993)}]{liu1993}%
  \BibitemOpen
  \bibfield  {author} {\bibinfo {author} {\bibfnamefont {W.}~\bibnamefont
  {Liu}}\ and\ \bibinfo {author} {\bibfnamefont {P.}~\bibnamefont {Tombesi}},\
  }\href@noop {} {\bibfield  {journal} {\bibinfo  {journal} {J. Mod. Opt.}\
  }\textbf {\bibinfo {volume} {5}},\ \bibinfo {pages} {181} (\bibinfo {year}
  {1993})}\BibitemShut {NoStop}%
\bibitem [{\citenamefont {Kral}(1990)}]{kral1990}%
  \BibitemOpen
  \bibfield  {author} {\bibinfo {author} {\bibfnamefont {P.}~\bibnamefont
  {Kral}},\ }\href@noop {} {\bibfield  {journal} {\bibinfo  {journal} {J. Mod.
  Opt.}\ }\textbf {\bibinfo {volume} {37}},\ \bibinfo {pages} {889} (\bibinfo
  {year} {1990})}\BibitemShut {NoStop}%
\bibitem [{\citenamefont {Satyanarayana}(1985)}]{PRD1985}%
  \BibitemOpen
  \bibfield  {author} {\bibinfo {author} {\bibfnamefont {M.~V.}\ \bibnamefont
  {Satyanarayana}},\ }\href {\doibase 10.1103/PhysRevD.32.400} {\bibfield
  {journal} {\bibinfo  {journal} {Phys. Rev. D}\ }\textbf {\bibinfo {volume}
  {32}},\ \bibinfo {pages} {400} (\bibinfo {year} {1985})}\BibitemShut
  {NoStop}%
\bibitem [{\citenamefont {Goldberg}(1989)}]{goldberg}%
  \BibitemOpen
  \bibfield  {author} {\bibinfo {author} {\bibfnamefont {D.~E.}\ \bibnamefont
  {Goldberg}},\ }\href@noop {} {\emph {\bibinfo {title} {Genetic Algorithms in
  Search, Optimization and Machine Learning}}}\ (\bibinfo  {publisher} {Boston,
  MA: Addison-Wesley},\ \bibinfo {year} {1989})\BibitemShut {NoStop}%
\end{thebibliography}
\end{document}